\documentclass[aps,pra,twocolumn,superscriptaddress]{revtex4-2} 
\usepackage{graphicx}
\usepackage{amsmath}
\usepackage{amssymb}
\usepackage{epstopdf}
\usepackage{amsfonts}
\usepackage{mathtools}
\usepackage{bigints}
\usepackage{xcolor}
\usepackage{braket}
\usepackage[normalem]{ulem}
\usepackage{booktabs}

\begin{document}
\title{Collective Superconducting-Qubit Interfaces for Microwave-to-Optical Quantum Transduction}
\author{Nicolas Dirnegger}
\address{Department of Electrical and Computer Engineering, UCLA, Los Angeles, CA, USA}
\author{Matthew Guibord}
\address{Department of Electrical and Computer Engineering, UCLA, Los Angeles, CA, USA}
\author{William Munizzi}
\address{Division of Physical Sciences, College of Letters and Science, University of California, Los Angeles (UCLA), Los Angeles, CA, USA}
\author{Chee Wei Wong}
\address{Department of Electrical and Computer Engineering, UCLA, Los Angeles, CA, USA}
\author{Prineha Narang}
\address{Department of Electrical and Computer Engineering, UCLA, Los Angeles, CA, USA}
\address{Division of Physical Sciences, College of Letters and Science, University of California, Los Angeles (UCLA), Los Angeles, CA, USA}

\date{\today}

\begin{abstract}
Microwave-to-optical quantum transduction is essential for connecting superconducting quantum processors through optical networks. Mechanically mediated transducers, however, remain limited by weak electromechanical coupling, dissipation, and added noise. In this work, we study a transducer in which an ensemble of superconducting qubits couples collectively to a shared optomechanical resonator. For weak signals, we show that ensemble size and optical pump strength together set the conversion efficiency, added noise, and bandwidth, with pump heating limiting efficiency to a finite optimal pump strength. Near half excitation, the collective coupling grows nearly linearly with ensemble size, accelerating transfer of an excitation into the mechanics. We show that this speedup is offset by relaxation of the additional excited qubits, so the single-qubit coupling and lifetime still set the coherence requirement. Transferring exactly one excitation also requires an anharmonic mechanical mode, which an auxiliary qubit coupled to the mechanics can provide. In a lossless prescribed-Kerr benchmark with four qubits and $g_{\rm sm}/(2\pi)=3.81\,\mathrm{kHz}$, a mechanical anharmonicity of $100\,\mathrm{kHz}$ raises the final joint single-photon intracavity target probability from $19.6\%$ to $99.7\%$. Finally, we quantify the effects of decoherence, disorder, residual qubit--qubit interactions, and pump-induced qubit noise, finding that pair-dependent interactions reduce the optical retrieval efficiency from 58\% to 35\%. These results establish the coherence and selectivity requirements for collective quantum transduction.
\end{abstract}
\maketitle

\section{Introduction}
\label{sec_introduction}

Distributed quantum computing and quantum networks require coherent interfaces between physically separated quantum systems~\cite{kimble2008quantum,Wehner2018QuantumInternetReview,Negrin,hughes2025compact,dirnegger2026distilled}. Superconducting circuits have emerged as a leading platform for quantum information processing~\cite{Krantz2019QubitGuide,Blais2021CircuitQED,arora2026multi}, and scaling these systems beyond a single cryostat requires interconnects between processors~\cite{ibm_united_states_2026_21997093}. Since superconducting qubits operate at microwave frequencies, their signals are strongly attenuated and susceptible to thermal fluctuations in room-temperature links. Converting microwave excitations into optical photons, which propagate with low loss in fiber, offers a path to networking superconducting processors~\cite{Lauk2020Perspectives,Lambert2020ConversionReview,Xu2024OptomechanicalTransducerReview,Sekine2026TransductionReview,Zhao2025PhotonicLinks,nongthombam2026fundamentals} and motivates performance metrics for coherent transduction~\cite{Zeuthen2020FiguresMerit,Wang2022QuantumCapacities,Zhong2022EntanglementTransduction}. Mechanically mediated transducers have made substantial experimental progress~\cite{Andrews2014Bidirectional,Forsch2020GroundStateConversion,Zhao2025QuantumEnabledNanomechanics}, including the conversion of superconducting qubit excitations into optical photons~\cite{Mirhosseini2020QubitOptical}.

Microwave-to-optical frequency conversion has been pursued through a variety of intermediary and nonlinear platforms, including ultracold atoms~\cite{Han2018Rydberg,Kumar2023NeutralAtoms}, spin ensembles~\cite{Li2017NVInterface,Liu2021NVConversion}, rare-earth-ion-doped crystals~\cite{Bartholomew2020YVO4,Rochman2023Erbium}, nanomechanical resonators~\cite{Forsch2020GroundStateConversion,Mirhosseini2020QubitOptical,Zhao2025QuantumEnabledNanomechanics,Yang2024MechanicalQubit}, and electro-optic materials~\cite{Lambert2020ConversionReview,Lauk2020Perspectives,Sekine2026TransductionReview,reineri2026microwave}. Optomechanical transducers utilize a mechanical resonator as an intermediary between superconducting microwave circuits and optical cavities, with the microwave--mechanical interaction mediated through piezoelectric or capacitive coupling and the mechanical--optical interaction mediated through radiation pressure~\cite{Aspelmeyer2014CavityOptomechanics,Han2020CavityPiezomechanics} (see Fig.~\ref{fig:generic_transducer} in Appendix~\ref{sec:background}). In optomechanical transducers, efficient and coherent conversion remains challenging since the electromechanical coupling rate competes with mechanical dissipation, thermally induced noise, and potential qubit decoherence~\cite{Zeuthen2020FiguresMerit,Mirhosseini2020QubitOptical,Zhao2025QuantumEnabledNanomechanics}. Increasing the electromechanical coupling without adding noise is therefore a central goal of transducer design.

Collective radiative enhancement enables broadband operation in weak-probe atomic quantum memories~\cite{rastogi2022superradiance}. Following this use of collective coherence to broaden the bandwidth of quantum-signal processing, we investigate a collective superconducting interface for mechanically mediated microwave-to-optical conversion. Ensembles of identical two-level systems coupled to a common mode exhibit collectively enhanced coupling, as described by the Tavis--Cummings and Dicke models~\cite{Dicke1954Coherence,Tavis1968NMolecule,Fink2009TavisCummings,Kubo2010SpinEnsemble,Amsuss2011CollectiveSpinStates}. In the symmetric Dicke manifold, the $(N+1)$-dimensional subspace of permutation-symmetric states with total spin $J=N/2$, the coupling on the transition between neighboring states containing $k$ and $k-1$ excitations is enhanced to $G_{\mathrm{sm}}(k)=g_{\mathrm{sm}}\sqrt{k(N-k+1)}$, where $N$ denotes the number of qubits and $g_{\mathrm{sm}}$ the single-qubit coupling strength. While the single-excitation ($k=1$) coupling scales as $\sqrt{N}$, the coupling on transitions near $k=N/2$ scales approximately linearly with $N$, enabling accelerated state transfer. Realizing this accelerated transfer requires the ensemble to remain collectively coherent and avoid unwanted transitions, despite dissipation and inhomogeneous broadening~\cite{Diniz2011CavityProtection,Kurucz2011Inhomogeneous,Mlynek2014DickeSuperradiance,MacQuarrie2015NVMechanical}.

\begin{figure*}[ht]
    \centering
    \includegraphics[width=\linewidth]{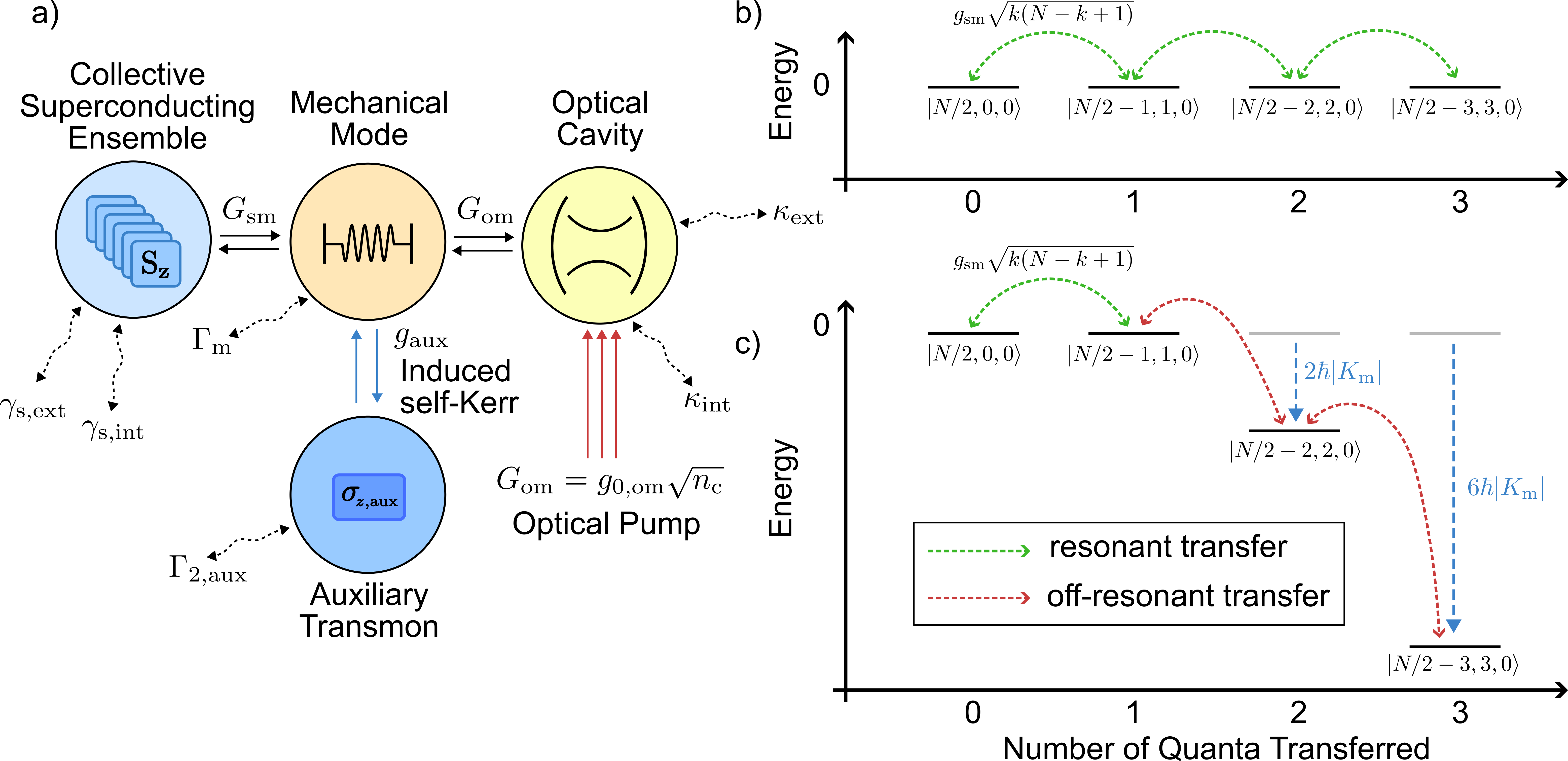}
    \caption{Collective superconducting--mechanical--optical transducer. \textbf{(a)} Three-mode architecture comprising the superconducting ensemble, mechanics, and optical cavity. The four-mode extension adds an auxiliary transmon. Couplings ($G_{\mathrm{sm}}$, $G_{\mathrm{om}}$, $g_{\mathrm{aux}}$), decay rates ($\gamma_s$, $\Gamma_m$, $\kappa$, $\Gamma_{2,\mathrm{aux}}$), external ports ($\gamma_{s,\mathrm{ext}}$, $\kappa_{\mathrm{ext}}$), and the optical pump are indicated. \textbf{(b,c)} Transfer ladders without and with auxiliary-induced anharmonicity. States are labeled $\ket{k,n_m,n_o}$ by the number of ensemble excitations and mechanical and optical quanta. In the anharmonic case, the first loading transition is tuned to resonance, while the $n$-phonon state is detuned by $-K_{\mathrm{m}}n(n-1)$.}
    \label{fig:setup}
\end{figure*}

Figure~\ref{fig:setup}(a) introduces our proposed architecture, with an ensemble of $N$ superconducting qubits coupled to a mechanical mode, which in turn is coupled to an optical cavity. An optional auxiliary transmon makes the mechanical mode anharmonic. Figures~\ref{fig:setup}(b) and \ref{fig:setup}(c) illustrate the resulting energy ladders without and with this anharmonicity.
This anharmonicity facilitates selective excitation transfer at higher Dicke excitation numbers by suppressing unwanted multiphonon transitions. In the weak-excitation regime, the linear calculation~\cite{Gardiner1985InputOutput,Clerk2010QuantumNoise,Aspelmeyer2014CavityOptomechanics} establishes the conversion efficiency, added noise, and spectral response of the collectively coupled interface. Beyond this regime, the nonlinear calculation examines excitation-selective transfer within the same architecture and identifies its sensitivity to local decoherence, thermalization, and disorder. We also compute the optical emission from a stored Dicke state in the presence of residual qubit--qubit interactions and pump-induced qubit noise. We find that the faster transfer near half excitation comes at the cost of more qubits excited during the transfer, and therefore more qubits that can decay. As a result, the coupling strength and lifetime of each individual qubit limit performance, and the mechanical nonlinearity must be large compared to the collective coupling for only one excitation to be transferred. Qubit--qubit interactions that vary from pair to pair break the permutation symmetry and cannot be removed by retuning.

\section{Dicke Ensemble Transducer}

We consider a mechanically mediated quantum transducer comprising an ensemble of $N$ superconducting qubits, a common mechanical mode, an optical cavity, and an optional auxiliary qubit as illustrated in Fig.~\ref{fig:setup}. The qubits couple to the mechanical mode, which exchanges excitations with the optical cavity through a pump-enhanced red-sideband interaction. We assume identical qubit transition frequencies and equal coupling strengths and phases unless disorder is explicitly introduced. Following the Tavis--Cummings model~\cite{Tavis1968NMolecule,Fink2009TavisCummings}, we define the collective superconducting operators as
\begin{equation}
\hat S_{\pm}=\sum_{j=1}^{N}\hat\sigma_{\pm}^{(j)},\qquad \hat S_z=\frac{1}{2}\sum_{j=1}^{N}\hat\sigma_z^{(j)},
\label{eq:collective_ops}
\end{equation}
where $\hat\sigma_+^{(j)}$ raises qubit $j$ from the ground state $\ket{0}$ to the excited state $\ket{1}$. The symmetric Dicke states~\cite{Dicke1954Coherence,Bartschi:2019zce,munizzi2024entropy,Keeler:2023xcx,Keeler:2023shl} are the permutation-symmetric states of an ensemble with $k$ excitations,
\begin{equation}\label{eq:dicke_state}
\ket{D_k^{N}} \equiv \ket{J=\frac{N}{2},\,m=k-\frac{N}{2}}, \qquad k=0,1,\dots,N.
\end{equation}
On the $N+1$ states in Eq.\ \eqref{eq:dicke_state}, the operator $\hat S_-$ acts as
\begin{equation}\label{eq:Sminus_action}
\hat S_-\ket{D_k^{N}} = \sqrt{k(N-k+1)}\,\ket{D_{k-1}^{N}},
\end{equation}
such that for an initially empty mechanical mode, a single-qubit coupling $g_{\mathrm{sm}}$ is enhanced in the $k \rightarrow k-1$ transition to
\begin{equation}\label{eq:Gsm_k}
G_{\mathrm{sm}}(k) = g_{\mathrm{sm}}\sqrt{k(N-k+1)}.
\end{equation}
If the mechanical mode holds $n$ phonons, this coupling gains a factor of $\sqrt{n+1}$, as shown in Appendix~\ref{sec:background}.

For the optical mode, we work in a frame rotating at the pump frequency $\omega_L$, with detuning $\Delta_o=\omega_L-\omega_o$. For a red-detuned pump, $\Delta_o\simeq-\omega_m$, and the linearized Hamiltonian is therefore
\begin{equation}
\hat H= \hat H_0+\hat H_{sm}+\hat H_{om},
\end{equation}
with the operators $\hat H_0$, $\hat H_{sm}$, and $\hat H_{om}$ given by
\begin{equation}\label{HamOps}
\begin{aligned}
\frac{\hat H_0}{\hbar} &= \omega_s\hat S_z +\omega_m\hat b^\dagger\hat b -\Delta_o\hat a^\dagger\hat a,\\
\frac{\hat H_{sm}}{\hbar} &= g_{sm}\left( \hat S_+\hat b+\hat S_-\hat b^\dagger \right),\\
\frac{\hat H_{om}}{\hbar} &= G_{om}\left(\hat a^\dagger\hat b+\hat a\hat b^\dagger\right).
\end{aligned}
\end{equation}
The operators $\hat a$ ($\hat a^\dagger$) and $\hat b$ ($\hat b^\dagger$) in Eq.\ \eqref{HamOps} are the annihilation and creation operators for the optical and mechanical modes, respectively. The coupling $G_{\mathrm{om}}=g_{0,\mathrm{om}}\sqrt{n_c}$ is the linearized optomechanical coupling, with $n_c$ the mean intracavity pump photon number. 

We consider the encoding of a signal qubit $\lvert\psi\rangle_q=\alpha\lvert0\rangle_q+\beta\lvert1\rangle_q$ into two neighboring symmetric Dicke states of an ensemble of $N$ superconducting qubits. For even $N$, we choose $\lvert0_L\rangle=\lvert D_{k-1}^{N}\rangle$ and $\lvert1_L\rangle=\lvert D_k^{N}\rangle$, with $k=N/2$. Starting with the ensemble in $\lvert D_{k-1}^{N}\rangle$, the coherent encoding operation is
\begin{equation}\label{eq:encoding}
\bigl(\alpha\lvert0\rangle_q+\beta\lvert1\rangle_q\bigr) \otimes\lvert D_{k-1}^{N}\rangle \rightarrow \lvert0\rangle_q\otimes \bigl(\alpha\lvert D_{k-1}^{N}\rangle +\beta\lvert D_k^{N}\rangle\bigr).
\end{equation}
When $k=N/2$, the encoded logical state lies near the center of the Dicke ladder, where $G_{\mathrm{sm}}(k)$ is largest. Realizing this encoding requires control of the chosen Dicke transition to suppress population transfer to neighboring Dicke states, while preserving the amplitudes $\alpha$, $\beta$ and their relative phase. For $k>1$, the lower logical state is not dark under $\hat S_-$, and instead couples to $\ket{D_{k-2}^{N},1_m}$ at the same frequency as the desired transition, such that mechanical anharmonicity cannot suppress it. Coherent conversion of an arbitrary superposition of adjacent Dicke states therefore requires additional transition-selective control. The nonlinear benchmarks quantify excitation selectivity and retrieval from prepared Dicke states. 

\subsection{Direct coupling to a common mechanical mode}
\label{sec:direct_collective_coupling}

A candidate implementation consists of electrically distinct superconducting transmons coupled through separate electrodes to one mechanical normal mode. An optical cavity samples the same mode in a spatially separated region. Electrostatic coupling between a transmon and silicon nanomechanics has been demonstrated \cite{bozkurt2025mechanical}, while a mechanically connected electrical--optical geometry has been demonstrated in silicon transduction \cite{Zhao2025PhotonicLinks, Bozkurt2023LongLivedPhonons}. Their extension to several independently controlled qubits is the proposed implementation considered here.

Let the displacement of the selected mechanical mode be $\mathbf u(\mathbf r)=x\mathbf f_m(\mathbf r)$, with
\begin{equation}
x=x_{\mathrm{zpf}}(\hat b+\hat b^\dagger),\quad x_{\mathrm{zpf}} = \sqrt{\frac{\hbar}{2m_{\mathrm{eff}}\omega_m}}.
\end{equation}
Each qubit couples through a displacement-dependent capacitance $C_{im}(x)=C_{im}(0)+C'_{im}x+\cdots$. In the independent-island approximation, expansion of the voltage-biased charging Hamiltonian (Appendix~\ref{app:direct_collective_interface}) gives the transition coupling
\begin{equation}
g_{im} = \frac{V_{\mathrm{dc},i}C'_{im}x_{\mathrm{zpf}}}{\hbar} \frac{2e\langle e_i|\hat n_i^{\mathrm{ch}}|g_i\rangle} {C_{\Sigma,i}},
\label{eq:direct_qubit_mechanical_coupling}
\end{equation}
up to a local phase convention. Here $C_{\Sigma,i}$ is the qubit's total capacitance, and $\hat n_i^{\mathrm{ch}}$ is the Cooper-pair number operator. The derivative $C'_{im}$ is evaluated along the same mechanical normal mode for every electrode. Projection onto the qubit transitions and the rotating-wave approximation give
\begin{equation}
\frac{\hat H_{\mathrm{sm}}}{\hbar} = \sum_{i=1}^{N} \left( g_{im}\hat\sigma_i^+\hat b + g_{im}^{*}\hat\sigma_i^-\hat b^\dagger \right).
\label{eq:direct_collective_hamiltonian}
\end{equation}
For equal, phase-aligned couplings $g_{im}=g_{\mathrm{sm}}$, this reduces to $H_{\mathrm{sm}}/\hbar =g_{\mathrm{sm}}(S_+b+S_-b^\dagger)$, and the collective enhancement follows from the Dicke matrix element rather than from an additional prefactor in the microscopic interaction.

For nonuniform couplings, the mechanically coupled single-excitation state is
\begin{equation}
|B\rangle = \frac{1}{G_B}\sum_i g_{im}|e_i\rangle, \quad G_B=\sqrt{\sum_i|g_{im}|^2},
\end{equation}
where $|e_i\rangle$ denotes the ensemble state with only qubit $i$ excited. The equal-weight Dicke state is recovered for uniform couplings. At higher excitation, nonuniform couplings generally do not preserve a closed symmetric Dicke ladder. The optical single-photon coupling is obtained from the same normalized mechanical mode through $g_{0,\mathrm{om}}=x_{\mathrm{zpf}}\,d\omega_o/dx$. 

\subsection{Auxiliary Qubit}

In order to make single-excitation transfer from the qubit ensemble into the mechanical resonator selective, we couple an auxiliary qubit to the mechanical resonator, which makes its spectrum anharmonic. We first analyze this interaction in the dispersive regime. The auxiliary qubit Hamiltonian and interaction with the mechanical resonator are defined as
\begin{equation}
    \frac{\hat{H}_{aux}}{\hbar} = \frac{\omega_{aux}}{2} \hat\sigma_z^{(\mathrm{aux})}, \quad
    \frac{\hat H_{am}}{\hbar}= g_{\rm aux}\left( \hat\sigma_+\hat b+\hat\sigma_-\hat b^\dagger \right).
\end{equation}
We define the auxiliary--mechanical detuning $\Delta_{\rm aux}=\omega_{\rm aux}-\omega_m$ and the small dispersive parameter $\lambda_{\rm aux}=\frac{g_{\rm aux}}{\Delta_{\rm aux}}$, requiring $\frac{2g_{\rm aux}\sqrt{n+1}}{|\Delta_{\rm aux}|}\ll1$ over the mechanically occupied subspace. In this limit, the auxiliary qubit acts as a dispersive control element. The expansion below defines $\chi_{\mathrm{aux}}$, $K_{\mathrm{m}}$, and $\alpha_m$, showing how this anharmonicity emerges. The explicit comparison parameters in Table~\ref{tab:model_parameters} lie outside this limit ($g_{\mathrm{aux}}/\Delta_{\mathrm{aux}}\approx0.51$), and we evaluate their anharmonicity using the Jaynes--Cummings spectrum (Sec.~\ref{sec:results}). Expanding the Jaynes--Cummings spectrum to fourth order in $g_{\rm aux}/\Delta_{\rm aux}$ gives
\begin{equation}
\begin{aligned}
\frac{\hat H_{am}^{\rm disp}}{\hbar} = (\omega_m+K_{\rm m})\hat n + \frac{\omega_{\rm aux}}{2}\hat\sigma_z +\\
\chi_{\rm aux} \left(\hat n+\frac{1}{2}\right)\hat\sigma_z + K_{\rm m}\hat n^2\hat\sigma_z
\end{aligned}
\end{equation}
where $\hat n=\hat b^\dagger\hat b$, as derived in Appendix~\ref{app:aux_disp}. The auxiliary-state-dependent dispersive pull of the mechanical resonance $\chi_{\rm aux}$ and the qubit-state-dependent mechanical Kerr nonlinearity $K_{\rm m}$ are defined as
\begin{equation}
\chi_{\rm aux}=g_{\rm aux} \lambda_{\rm aux} + K_{\rm m},
\qquad
K_{\rm m} = -g_{\rm aux} \lambda_{\rm aux}^3.
\end{equation}
Combining the transduction couplings with the dispersive auxiliary sector gives
\begin{align}
\frac{\hat H_{\rm disp}}{\hbar} ={}& \omega_s\hat S_z -\Delta_o\hat a^\dagger\hat a + (\omega_m+K_{\rm m})\hat n + \frac{\omega_{\rm aux}}{2}\hat\sigma_z \nonumber\\
&+ g_{sm}\left( \hat S_+\hat b+\hat S_-\hat b^\dagger \right) + G_{om}\left( \hat a^\dagger\hat b+\hat a\hat b^\dagger \right)
\nonumber\\
&+ \chi_{\rm aux} \left(\hat n+\frac{1}{2}\right)\hat\sigma_z + K_{\rm m}\hat n^2\hat\sigma_z .
\end{align}
For the auxiliary qubit in its ground state, $\hat\sigma_z\rightarrow-1$, the frequency of the mechanical transition from $n$ to $n+1$ phonons is
\begin{equation}
\omega_{n,n+1}^{(g)} = \omega_m-\chi_{\rm aux}-2nK_{\rm m}.
\end{equation}
The first transition is thereby shifted by $-\chi_{\rm aux}$, and successive transitions are separated by the anharmonicity
\begin{equation}
\alpha_m = \omega_{n+1,n+2}^{(g)} - \omega_{n,n+1}^{(g)} = -2K_{\rm m},
\end{equation}
which is positive for $\Delta_{\rm aux}>0$. The first Dicke transition is tuned to the shifted frequency $\omega_{0,1}^{(g)}$ (Appendix~\ref{app:rotating_frame}). 

\subsection{Input--output response}

\label{sec:io_compact}
To characterize microwave-to-optical quantum conversion, we use the passive red-sideband input--output description in the weak-excitation regime via the Heisenberg--Langevin formalism. The collective superconducting ensemble is represented by the normalized bright-mode operator $\hat s=\hat S_-/\sqrt{N}$, for which $[\hat s,\hat s^\dagger]\simeq1$ near the collective ground state, and the superconducting ensemble--mechanical coupling is $G_{\mathrm{sm}}\equiv G_{\mathrm{sm}}(1)=g_{\mathrm{sm}}\sqrt{N}$, the $k=1$ value of Eq.\ \eqref{eq:Gsm_k}. The input and output fields remain quantum operators. In frequency space we write $v_{\rm out}(\omega) = \Xi(\omega) v_{\rm in}(\omega)$, where
\begin{equation}
    \Xi(\omega) = C \left(-i\omega I-A\right)^{-1} B+ D, 
\label{eq:Xi_general_main}
\end{equation}
is the general response function of the input-output relation.  The full state-space matrices for the three- and four-mode architectures are given in Appendix~\ref{app:io_matrices}. For the three-mode Dicke--mechanical--optical transducer, the forward superconducting ensemble-to-optical conversion amplitude is
\begin{equation}
\Xi_{as}^{(3)}(\omega)
=
\frac{
\sqrt{\gamma_{s,\mathrm{ext}}\kappa_{\mathrm{ext}}}\,
G_{sm}G_{om}
}{
D_3(\omega)
},
\label{eq:Xi3_main}
\end{equation}
with $\kappa_{ext}$ the external optical coupling rate and $\gamma_{s,ext}$ the external coupling rate of the ensemble. Furthermore, the denominator $D_3(\omega)$ is
\begin{equation}
D_3(\omega) = \chi_s^{-1}(\omega)\chi_m^{-1} (\omega)\chi_a^{-1}(\omega) + G_{sm}^2\chi_a^{-1} (\omega) + G_{om}^2\chi_s^{-1}(\omega).
\label{eq:D3_main}
\end{equation}
The conversion efficiency is $\eta_3(\omega)=\left|\Xi_{as}^{(3)}(\omega)\right|^2.$ On resonance this reduces to
\begin{equation}
\eta_3(0)
=
\eta_s\eta_o
\frac{4C_{sm}C_{om}}
{\left(1+C_{sm}+C_{om}\right)^2},
\label{eq:eta3_main}
\end{equation}
where $\eta_s=\frac{\gamma_{s,\mathrm{ext}}}{\gamma_s}$, $\eta_o=\frac{\kappa_{\mathrm{ext}}}{\kappa}$, $C_{sm}=\frac{4G_{sm}^2}{\gamma_s\Gamma_m}$, and $C_{om}=\frac{4G_{om}^2}{\kappa\Gamma_m}$  are the extraction efficiencies and cooperativities on resonance. For the four-mode system, retaining the auxiliary system in the weak-excitation limit gives
\begin{equation}
\Xi_{as}^{(4)}(\omega) = \frac{ \sqrt{\gamma_{s,\mathrm{ext}}\kappa_{\mathrm{ext}}}\, G_{sm}G_{om}\chi_q^{-1}(\omega)}{D_4(\omega)},
\label{eq:Xi4_main}
\end{equation}
where $\chi_q$ is the auxiliary susceptibility defined in Appendix~\ref{app:efficiency_bandwidth}, and 
\begin{equation}
D_4(\omega)
=
\chi_q^{-1}(\omega)D_3(\omega)
+
g_{\mathrm{aux}}^2
\chi_s^{-1}(\omega)\chi_a^{-1}(\omega).
\label{eq:D4_main}
\end{equation}
Equivalently, the auxiliary mode dresses the mechanical susceptibility as $\chi_{m,\mathrm{eff}}^{-1}(\omega) = \chi_m^{-1}(\omega) + g_{\mathrm{aux}}^2\chi_q(\omega).$ The reported conversion bandwidth is the width of the connected half-maximum interval containing the selected conversion peak. For half-maximum crossings $\omega_-$ and $\omega_+$, $B=(\omega_+-\omega_-)/(2\pi)$. Disconnected peaks are treated as separate conversion bands. In the weak-coupling, mechanically limited regime,
\begin{equation}
\Gamma_{\mathrm{tot}}^{(3)}
\simeq
\Gamma_m+\Gamma_{sm}+\Gamma_{om},
\label{eq:Gammaeff3_main}
\end{equation}
where $\Gamma_{sm}=\frac{4G_{sm}^2}{\gamma_s}$ and  $\Gamma_{om}=\frac{4G_{om}^2}{\kappa}$ so that $B_3\simeq\Gamma_{\mathrm{tot}}^{(3)}/2\pi$. The corresponding four-mode expression contains the additional auxiliary-induced mechanical damping derived in Appendix~\ref{app:efficiency_bandwidth}. The input-referred added noise is obtained directly from the same scattering matrix,
\begin{equation}
n_{\mathrm{add}}(\omega)
=
\frac{
\sum_{j\neq s_{\mathrm{ext}}}
\left|\Xi_{a,j}(\omega)\right|^2 n_j
}{
\left|\Xi_{a,s_{\mathrm{ext}}}(\omega)\right|^2
}.
\label{eq:nadd_main}
\end{equation}
The complete thermal-occupancy and added-noise derivations are given in Appendix~\ref{app:thermal_occupancy}. 

\begin{figure*}[ht]
    \centering
    \includegraphics[width=0.75\linewidth]{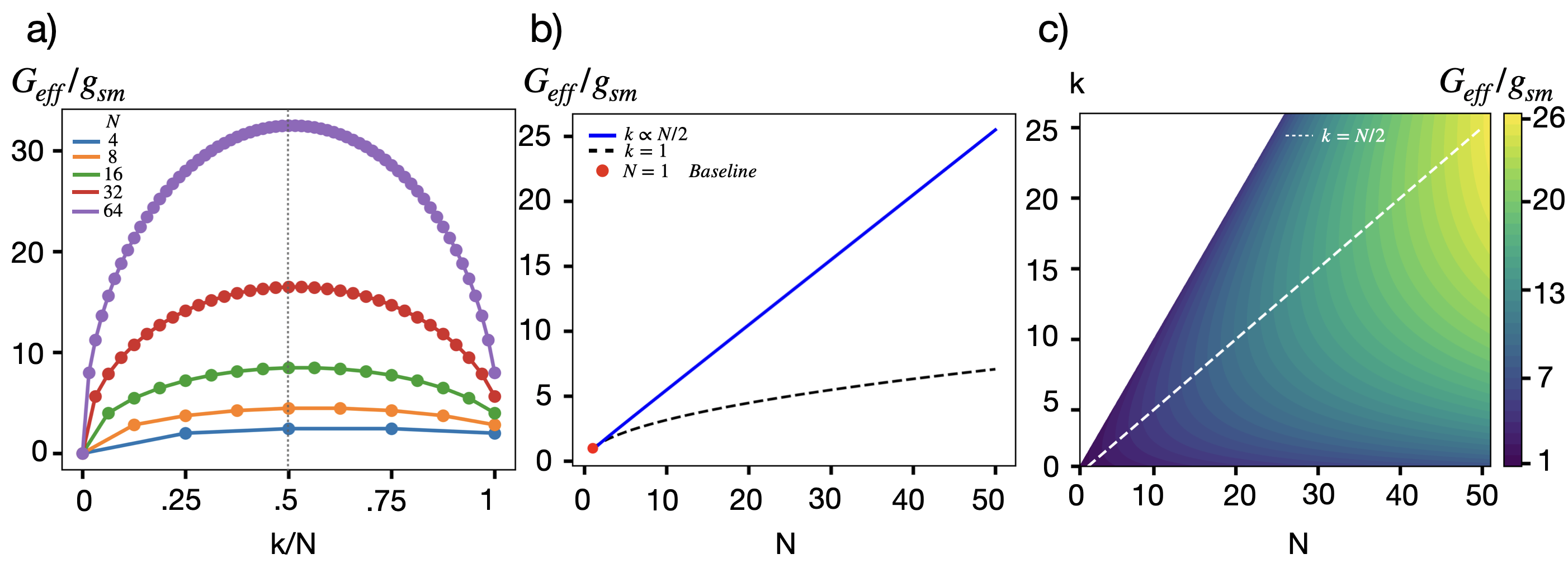}
    \includegraphics[width=0.75\linewidth]{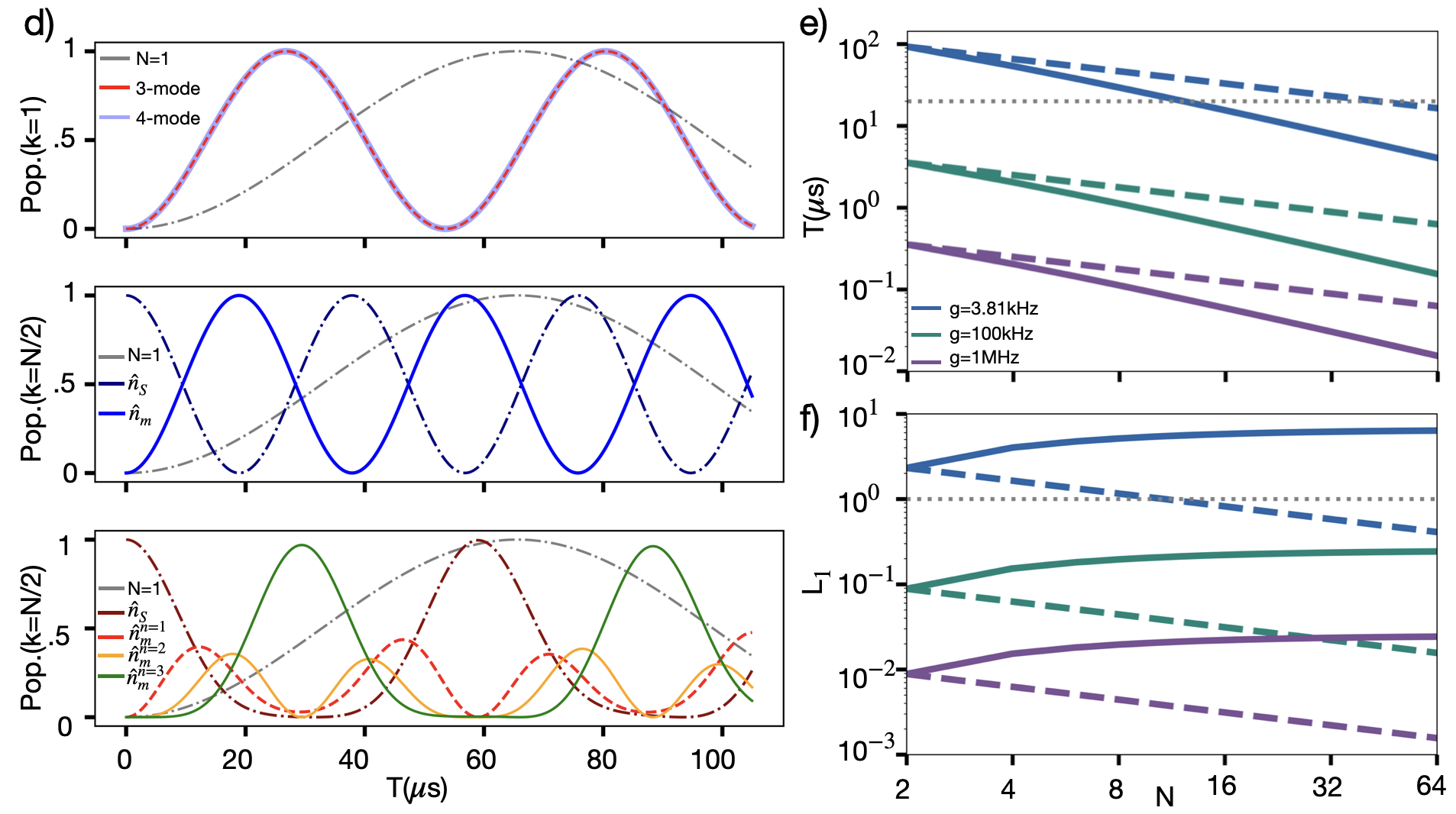}
    \caption{Collective enhancement, exchange dynamics, and relaxation exposure.
    \textbf{(a)--(c)} Normalized coupling $G_{\mathrm{sm}}(k)/g_{\mathrm{sm}}=\sqrt{k(N-k+1)}$: \textbf{(a)} dependence on $k/N$ for $N=4,8,16,32,64$, with markers at integer $k$. \textbf{(b)} comparison of $k=N/2$ (solid) and $k=1$ (dashed), with the single-qubit reference marked in red. \textbf{(c)} dependence on $N$ and $k$, with $k>N$ excluded. 
    \textbf{(d)} Population dynamics for $k=1$ (top) and $k=N/2$ (middle and bottom), including the three-/four-mode comparison and state-resolved channels indicated in the legends. Gray dash-dotted curves show the single-qubit reference.
    \textbf{(e)} Ideal selective-loading duration $T_{\mathrm{load}}=\pi/G_{\mathrm{sm}}(k)$ for a resonant $\sin^2$ pulse of peak amplitude $g_{\mathrm{sm}}$.
    \textbf{(f)} Integrated local-relaxation exposure $\mathcal L_1$. In (e,f), solid/dashed curves denote $k=N/2$/$k=1$ where blue, teal, and purple denote $g_{\mathrm{sm}}/(2\pi)=3.81~\mathrm{kHz}, 100~\mathrm{kHz},1~\mathrm{MHz}$. Dotted lines indicate $T_1=20~\mu\mathrm{s}$ in (e) and $\mathcal L_1=1$ in (f).}
    \label{fig:state_dynamics}
    \label{fig:Effective_Coupling_Dynamics}
\end{figure*}

\subsection{Master Equation}
To investigate collective state-transfer dynamics beyond the weak-excitation regime, we solve the time-dependent Lindblad master equation for the coupled system~\cite{Lindblad1976,Breuer2002OpenQuantum,Johansson2013QuTiP2}. Near the half-excited Dicke manifold, \(k=N/2\), the Holstein--Primakoff approximation is no longer sufficient to describe the excitation-dependent collective interaction. We therefore retain the full Dicke ladder, including the possibility of transitions into unwanted excitation manifolds. 
The open-system dynamics are governed by

\begin{equation}
\frac{d\hat\rho}{dt}
=-\frac{i}{\hbar}[\hat H(t),\hat\rho]
+\sum_\mu\mathcal D[\hat L_\mu]\hat\rho,
\end{equation}

where the Lindblad dissipator is defined as $\mathcal D[\hat L]\hat\rho = \hat L\hat\rho\hat L^\dagger -\frac12\left\{\hat L^\dagger\hat L,\hat\rho\right\}$. The nonlinear benchmark in Fig.~\ref{fig:auxiliary_loss} uses sequential mechanical loading and optical readout, with the superconducting ensemble--mechanical interaction suppressed during readout. Figure~\ref{fig:linear} instead uses the steady-state response with both interactions on.

In addition to thermalization through the cold environment, we include a phenomenological mechanical bath generated by absorption of the optical pump~\cite{Zhao2025QuantumEnabledNanomechanics,Meenehan2015PulsedHeating}. The cold and optically induced baths are characterized by coupling rates $\Gamma_f$ and $\Gamma_h(t)$ and occupations $\bar n_f$ and $\bar n_h(t)$, respectively. Assuming independent, locally Markovian reservoirs with approximately flat spectra over the relevant mechanical transitions, their contribution to the master equation is
\begin{equation}
\begin{aligned}
\mathcal{L}_m(t)\rho ={}& \sum_{\nu=f,h} \Gamma_\nu(t)\bigl[\bar n_\nu(t)+1\bigr] \mathcal{D}[\hat b]\rho\\
&+
\sum_{\nu=f,h} \Gamma_\nu(t)\bar n_\nu(t) \mathcal{D}[\hat b^\dagger]\rho .
\end{aligned}
\end{equation}
The corresponding intrinsic mechanical damping and thermal injection rate are
\begin{align}
\Gamma_m(t)
&=\Gamma_f+\Gamma_h(t),\\
Q_m(t)
&=\Gamma_f\bar n_f+\Gamma_h(t)\bar n_h(t).
\end{align}
For stationary pumping, these quantities define the effective bath occupation $\bar n_m=Q_m/\Gamma_m$. This bath occupation is distinct from the mechanical-mode population, which also depends on coherent exchange and optical extraction. The pump-dependent bath parameters are device-specific inputs. For pulsed operation their evolution can additionally account for finite heating and cooling times. Cavity-induced mechanical damping is not added separately when the optical mode and its interaction are retained explicitly. We adopt a phenomenological hot-bath model with $\Gamma_h/(2\pi)=(1.1\,\mathrm{kHz})n_c^{0.61}$ and $\bar n_h=1.1n_c^{0.30}$, motivated by measurements in silicon optomechanical crystals~\cite{ren2020two}. The coefficients are used as a reference heating scenario. Extension of the damping power law below its principal fitting range is treated as a model assumption. We track subsystem occupations, target-state populations, and optical emission.

For identical qubits and uniform couplings, the collective Hamiltonian preserves total spin. Independent local relaxation and dephasing can nevertheless transfer population out of the initial $J=N/2$ manifold. For larger ensembles, we use a permutation-invariant Liouville-space representation retaining the total-spin sectors generated by identical local relaxation and dephasing~\cite{shammah2018open}. For small ensembles, we retain the complete $2^N$ qubit Hilbert space and apply independent $T_1$ and $T_\phi$ collapse operators to each transmon. The two representations are benchmarked against one another for ensembles up to $N=6$. Static qubit-frequency disorder is treated using the full individual-qubit representation, because it breaks permutation symmetry.

To model the effect of optical pumping on the qubits, we add pump-dependent relaxation, excitation, and pure-dephasing rates to the intrinsic qubit dissipation. These rates act through the same local operators as the intrinsic channels and scale with the pump occupation $n_c$ (Appendix~\ref{app:finite_gate_retrieval}). Because they are device dependent, we treat them as sensitivity parameters. Residual qubit--qubit interactions are included through $H_{\mathrm{res}}/\hbar=\sum_{i<j}\zeta_{ij}\hat n_i\hat n_j$. We treat uniform couplings, which shift the collective transitions, separately from pair-dependent couplings, which break permutation symmetry (Appendix~\ref{app:zz_model}).

The collapse operators are given in Appendix~\ref{app:master_equation}. Model parameters, their provenance, numerical settings, and the dissipative channels retained in each calculation are specified in Appendix~\ref{app:parameters}, Table~\ref{tab:model_parameters}.

\section{Simulation Results}\label{sec:results}

\begin{figure*}
    \centering
    \includegraphics[width=0.75\linewidth]{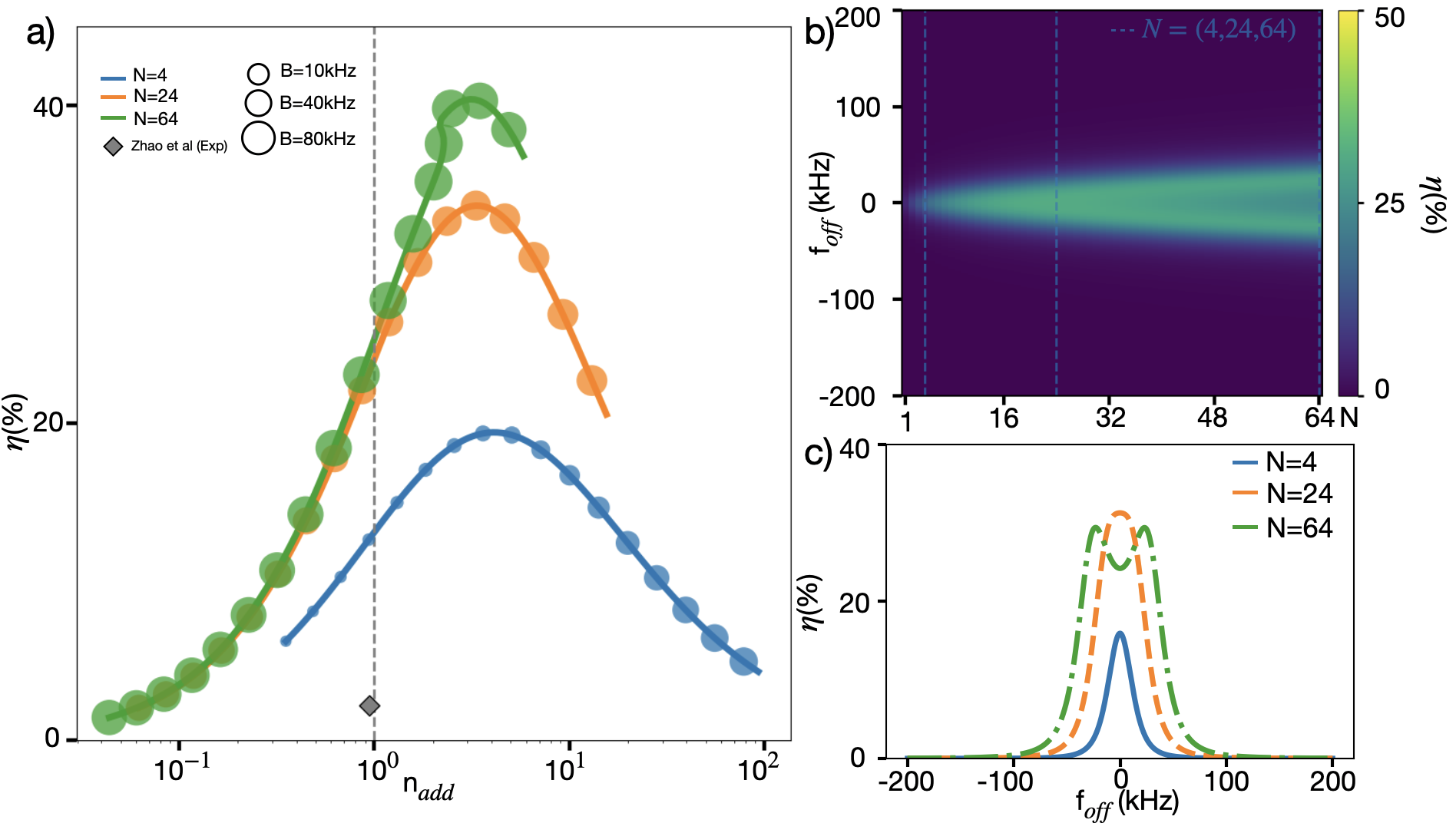}
    \caption{Linear quantum-transduction performance and conversion spectra. \textbf{(a)} External efficiency $\eta$ versus input-referred added noise for $N=4,24,64$ under an optical-pump sweep, including the assumed pump-dependent mechanical heating. Each efficiency--noise pair is evaluated at the same conversion-peak frequency. Filled-marker areas encode the connected $3$-dB bandwidth and hollow circles provide the size key. The dashed line marks $n_{\mathrm{add}}=1$, and the gray diamond shows the experimental reference $(\eta,n_{\mathrm{add}})=(2.2\%,0.94)$ \cite{Zhao2025QuantumEnabledNanomechanics}. \textbf{(b)} Absolute efficiency $\eta(f_{\mathrm{offset}},N)$ at fixed pump $n_c = 100$ and bath settings where dashed lines mark $N=4,24,64$. \textbf{(c)} Representative spectra for these ensemble sizes, illustrating the evolution from a narrow central resonance to a broadened response and in strong-coupling regime a two-peak spectrum.}
    \label{fig:linear}
\end{figure*}

Figure~\ref{fig:Effective_Coupling_Dynamics} quantifies the collective enhancement of the superconducting ensemble–mechanical interaction across the symmetric Dicke ladder. For an ensemble of \(N\) identically coupled qubits, the transition \(\lvert D_k^N,0_m\rangle\rightarrow\lvert D_{k-1}^{N},1_m\rangle\) has the matrix element $G_{sm}(k)$ from Eq.\ \eqref{eq:Gsm_k}, with \(g_{\mathrm{sm}}\) the microscopic single-qubit coupling. The enhancement is therefore determined by both the ensemble size and its excitation number. 

At the lowest transition, \(G_{\mathrm{sm}}(1)=g_{\mathrm{sm}}\sqrt N\), whereas near the middle of the ladder the coupling approaches \(g_{\mathrm{sm}}N/2\) for large \(N\). The largest matrix elements consequently occur near half excitation, providing a stronger dependence on ensemble size than is available in the single-excitation regime. Whether this speeds up transfer depends on selecting the desired transition and keeping the ensemble collectively coherent.

\begin{figure*}
    \centering
    \includegraphics[width=0.75\linewidth]{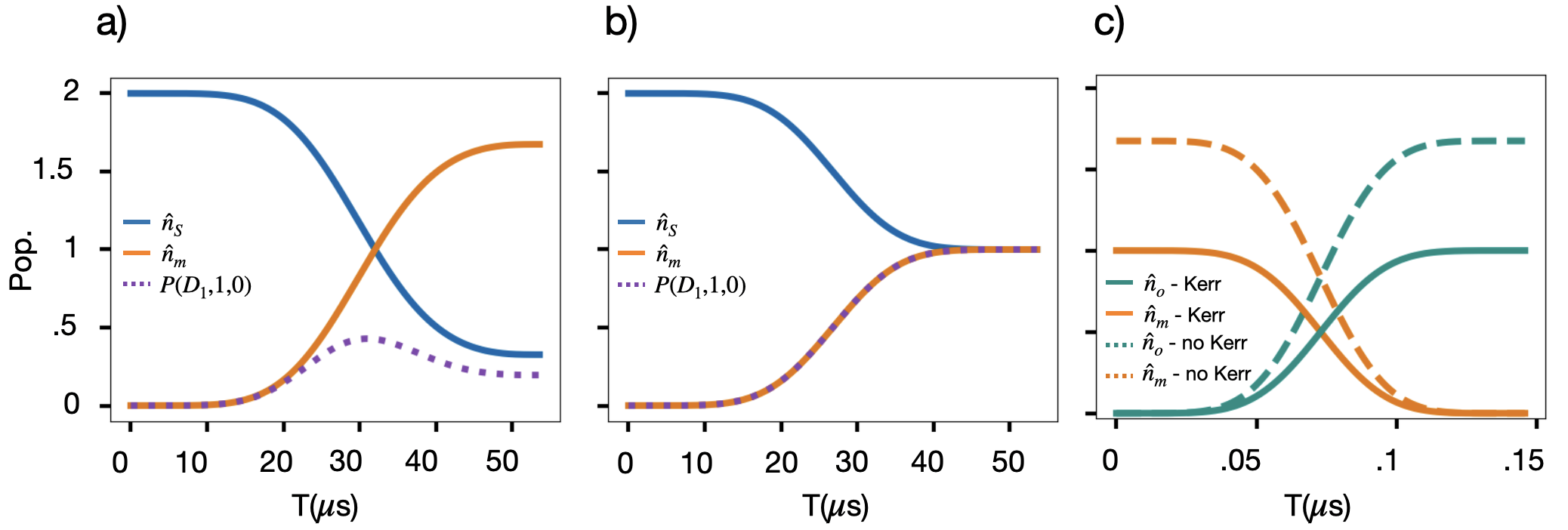}
    \includegraphics[width=0.75\linewidth]{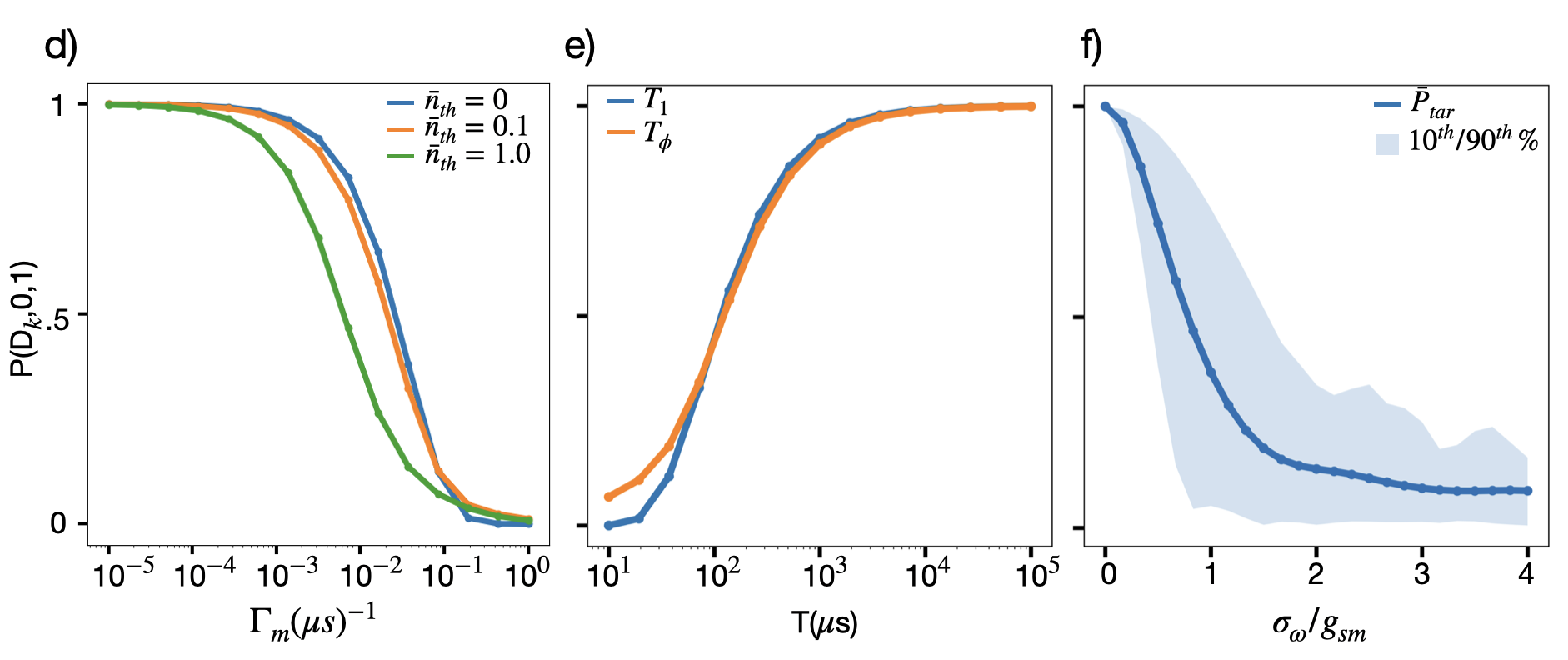}
    \caption{Kerr-assisted selective transfer and sensitivity to imperfections. The $N=4$ ensemble starts in $\lvert D_2^{4},0_m,0_o\rangle$. The Kerr-on calculations use the prescribed anharmonicity $\alpha_m/(2\pi)=100~\mathrm{kHz}$, corresponding to $K_m/(2\pi)=-50~\mathrm{kHz}$ under $\alpha_m=-2K_m$, with the first mechanical transition tuned to resonance. Sequential, nonoverlapping $\sin^2$ coupling pulses have peak amplitudes $g_{\mathrm{sm}}/(2\pi)=3.81~\mathrm{kHz}$ and $G_{\mathrm{om}}/(2\pi)=3.43~\mathrm{MHz}$.
    \textbf{(a,b)} Lossless mechanical loading without and with a prescribed Kerr nonlinearity. Blue and orange curves show mean ensemble and mechanical occupations, respectively, while purple dotted curves show $P(D_1^{4},1_m,0_o)$.
    \textbf{(c)} Subsequent lossless optical readout with superconducting ensemble--mechanical exchange switched off. Orange/teal curves show mechanical/optical occupations and solid/dashed curves correspond to Kerr-on/Kerr-off loading.
    \textbf{(d)--(f)} Final target probability $P_{\mathrm{tar}}=P(D_1^{4},0_m,1_o)$ versus \textbf{(d)} mechanical damping at $\bar n_{\mathrm{th}}=0,0.1,1$, \textbf{(e)} independently varied $T_1$ and $T_\phi$, and \textbf{(f)} static frequency-disorder width $\sigma_\omega/g_{\mathrm{sm}}$. The disorder curve and shading give the mean and $10^{th}$--$90^{th}$\% across realizations.}
    \label{fig:auxiliary_loss}
\end{figure*}

Figure~\ref{fig:state_dynamics} also connects the collective matrix element to the dynamics of mechanical loading. The desired operation transfers one excitation from the ensemble into the mechanical mode while retaining the remaining \(k-1\) excitations in the Dicke state. For an isolated resonant transition driven by a pulse \(g_{\mathrm{sm}}(t)=g_{\mathrm{sm}}\sin^2(\pi t/T_{\mathrm{load}})\), complete loading occurs at \(T_{\mathrm{load}}=\pi/G_{sm}(k)\). The ideal duration therefore decreases as \(N^{-1/2}\) for \(k=1\) and approximately as \(N^{-1}\) near half excitation. Applying it to the full system requires sufficient suppression of neighboring transitions, whose excitation would redistribute population beyond the intended transfer channel. The Fig.~\ref{fig:Effective_Coupling_Dynamics}(d) panels illustrate the associated exchange dynamics relative to the single-qubit reference. In the displayed single-excitation example, the three- and four-mode population traces nearly overlap and evolve faster than the $N=1$ reference. The half-excited examples show faster population exchange and, in the state-resolved traces, occupation of higher mechanical excitation states. These oscillations show reversible exchange, not selective single-excitation loading, which we examine in Fig.~\ref{fig:auxiliary_loss}.

The reduction in loading time must be assessed together with the number of excitations exposed to independent qubit relaxation. Along the ideal selective-loading trajectory, the integrated relaxation exposure is
\begin{equation}
\mathcal L_1
=\frac{1}{T_1}\int_0^{T_{\mathrm{load}}}
\langle \hat N_q(t)\rangle_{\mathrm{ideal}}\,dt
=\left(k-\frac12\right)\frac{T_{\mathrm{load}}}{T_1},
\end{equation}
where \(T_1\) is the single-qubit relaxation time. This quantity measures the importance of relaxation during the pulse, not the transfer probability. For \(k=N/2\), the increasing population of excited qubits offsets the shorter loading duration, giving \(\mathcal L_1\rightarrow\pi/(g_{\mathrm{sm}}T_1)\) at large \(N\). A larger ensemble therefore speeds up the exchange but does not reduce exposure to relaxation, so \(g_{\mathrm{sm}}T_1\) remains the key requirement.

Figure~\ref{fig:linear} connects this collective interaction to conventional transduction observables in the weak-excitation regime. Here the ensemble is represented by its linearized bright mode, with $G_{\mathrm{sm}}=G_{\mathrm{sm}}(1)$, and the complete input–output response is used instead of the nonlinear Dicke ladder. Figure~\ref{fig:linear}(a) shows the external conversion efficiency against input-referred added noise while the intracavity optical pump occupation is varied. Increasing the pump strengthens the optomechanical interaction and can initially improve conversion, but the accompanying absorption-induced mechanical bath produces an increase in thermal-noise injection. Efficiency therefore peaks at a finite pump power. Over the displayed pump sweeps, the maximum efficiencies are approximately $20\%$, $30\%$, and $40\%$ for $N=4,24,64$, respectively. At comparable added-noise levels below one photon, the larger ensembles also reach higher efficiencies than the $N=4$ case. The efficiency maxima, however, lie above $n_{\mathrm{add}}=1$, so the pump setting that maximizes efficiency is not the one that keeps $n_{\mathrm{add}}$ below one. The marker areas provide the corresponding connected $3$-dB bandwidths. 

Figure~\ref{fig:linear}(b) and ~\ref{fig:linear}(c) show how increasing the collective superconducting ensemble–mechanical coupling modifies the conversion spectrum. At weak collective coupling the response contains a narrow resonance whose linewidth is predominantly set by the induced mechanical damping rates. Increasing $N$ broadens the useful response until coherent hybridization becomes appreciable, at which point the spectrum develops two resolved peaks. In this regime we take the bandwidth from the full spectrum, not from a single mechanical linewidth. This splitting arises within the linear model and does not indicate a breakdown of the weak-excitation approximation. Ensemble size and pump strength therefore act as two separate controls on the operating point. Increasing collective coupling changes the ensemble--mechanical cooperativity without changing the microscopic single-qubit coupling, while the optical pump controls the optomechanical interaction and the assumed absorption heating.

The nonlinear selective-transfer problem is addressed in Fig.~\ref{fig:auxiliary_loss}. Unlike the linear response of Fig.~\ref{fig:linear}, the calculation retains the excitation dependence of the Dicke ladder and distinguishes the desired joint state from the total subsystem occupations. Without mechanical anharmonicity, successive Dicke–mechanical transitions remain nearly resonant and population spreads into neighboring excitation manifolds. Introducing a prescribed mechanical Kerr nonlinearity suppresses this multiphonon loading and enables substantially more selective removal of a single excitation from the initially half-excited ensemble (Fig.~\ref{fig:auxiliary_loss}(a)--(b)). The Kerr-off sequence produces a larger mean optical occupation since its preceding loading stage populated more mechanical excitations (Fig.~\ref{fig:auxiliary_loss}(c)). Subsequent suppression of the superconducting ensemble–mechanical interaction allows the mechanical excitation to be transferred into the optical mode (Fig.~\ref{fig:auxiliary_loss}(c)). The anharmonicity dependence of this selective-transfer mechanism is quantified in Fig.~\ref{fig:alpha} in Appendix \ref{app:aux_disp}. With the loading and readout pulses held fixed, a prescribed anharmonicity $\alpha_m/(2\pi)=100~\mathrm{kHz}$ yields a final optical target-state probability of $99.71\%$, while strongly suppressing optical multiphoton occupation. The sweep thus sets the anharmonicity needed for near-unity selective transfer in this lossless benchmark. For a microscopic spectral comparison, a two-level auxiliary qubit with $g_{\mathrm{aux}}/(2\pi)=1~\mathrm{MHz}$ and $\Delta_{\mathrm{aux}}/(2\pi)=1.9565~\mathrm{MHz}$ produces approximately the same $100~\mathrm{kHz}$ anharmonicity in the exact Jaynes--Cummings spectrum. This parameter pair is outside the controlled far-dispersive limit and is not used to derive the prescribed Kerr coefficient through a fourth-order expansion.

\begin{figure*}
    \centering
    \includegraphics[width=\linewidth]{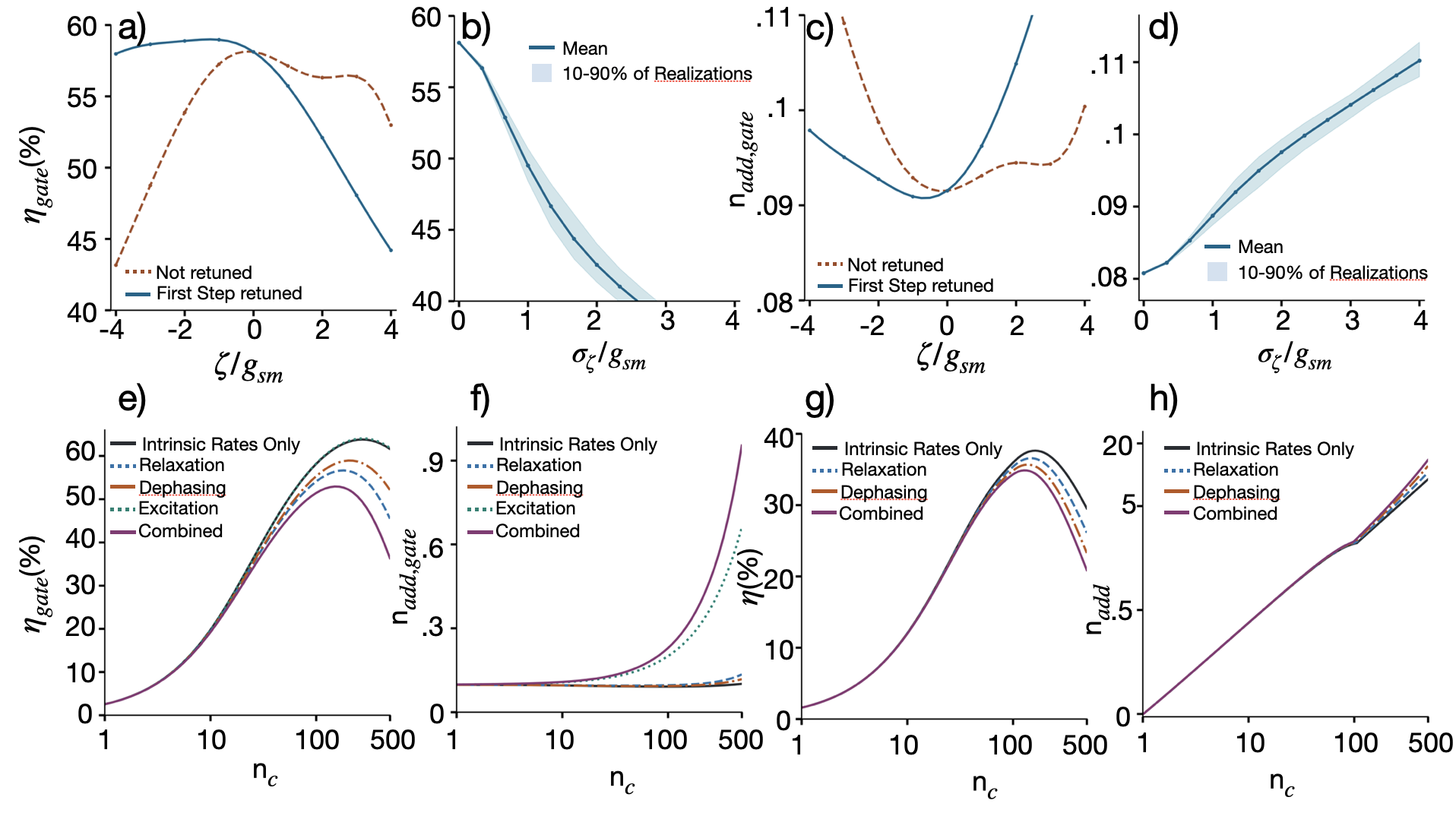}
    \caption{(a) External retrieval efficiency under uniform ZZ coupling for an initially prepared Dicke state $\lvert D_3^{6}\rangle$, with $g_{\rm sm}/2\pi=10\,\mathrm{kHz}$ and pump occupation $n_c=100$. Blue solid and orange dotted curves show the original frequency setting and retuning of the first emission transition, respectively. (b) External retrieval efficiency versus RMS pairwise ZZ disorder $\sigma_\zeta/g_{\rm sm}$. At each nonzero disorder strength, 24 Gaussian pair-coupling patterns are centered to zero pair mean and normalized to the specified RMS. Lines give the mean and shading gives the 10th-90th percentiles. (c) Input-referred added noise under uniform ZZ coupling. (d) Input-referred added noise versus RMS pairwise ZZ disorder. (e) External retrieval efficiency versus pump occupation $n_c$. Charcoal includes intrinsic qubit rates while blue dashed, orange dash-dotted, and teal dotted curves additionally include pump-induced relaxation, pure dephasing, and excitation, respectively. Purple includes all three added channels. The prescribed added rates are $(\Gamma_\downarrow,\Gamma_\phi,\Gamma_\uparrow)=(0.005,0.005,0.0005)(n_c/100)\,\mu\mathrm{s}^{-1}$. Intrinsic lifetimes are $T_1=100\,\mu\mathrm{s}$ and $T_\phi=200\,\mu\mathrm{s}$. In (e) and (f), the mechanical bath occupation is $0.1$ with no pump heating; (g) and (h) include the pump-dependent mechanical bath of Eq.\ \eqref{eq:hot_bath_power_laws}. (f) Input-referred added noise versus pump occupation. (g) Peak external conversion efficiency in the continuous weak-probe model. (h) Input-referred spectral added noise in continuous conversion.} 
    \label{fig:robustness}
\end{figure*}

Figure~\ref{fig:auxiliary_loss}(d)--(f) quantify the sensitivity of this Kerr-assisted protocol to the principal decoherence mechanisms. Increasing mechanical damping reduces the final target-state population, with thermal occupation further degrading the transfer through phonon injection. Independent qubit relaxation and pure dephasing become increasingly detrimental when their characteristic times approach the duration of the loading and readout sequence. Static qubit-frequency disorder similarly destroys permutation symmetry and reduces the probability of remaining in the intended collective state. Reaching the target state therefore requires the ensemble to stay collectively coherent, not just to give up an excitation.\\

Figure~\ref{fig:robustness} examines the sensitivity of transduction to residual ZZ interactions and pump-induced qubit noise at $g_{\rm sm}/2\pi=10\,\mathrm{kHz}$. Unlike Fig.~\ref{fig:auxiliary_loss}, these calculations use the explicit auxiliary qubit of Table~\ref{tab:model_parameters} instead of a prescribed Kerr term, and keep the ensemble--mechanical and optomechanical couplings on throughout the collection window, with the optical cavity adiabatically eliminated (Appendix~\ref{app:finite_gate_retrieval}). For retrieval from the initially prepared Dicke state $\lvert D_3^{6}\rangle$, we evaluate the external optical counts
\begin{equation}
    C_j =
    \int_0^{T_{\rm g}}
    \left\langle
        L_{\rm out}^{\dagger}L_{\rm out}
    \right\rangle_j dt,
    \qquad T_{\rm g}=150\,\mu\mathrm{s},
\end{equation}
and define
\begin{equation}\label{Eq32}
    \eta_{\rm gate}=\frac{C_3-C_0}{3},
    \qquad
    n_{\rm add,gate}=\frac{C_0}{\eta_{\rm gate}}.
\end{equation}
Here $C_0$ accounts for photons emitted with the ensemble initially in its ground state, and dividing by the three stored excitations gives the retrieval efficiency per excitation. At zero ZZ coupling and $n_c=100$, the model gives $\eta_{\rm gate}=58.1\%$ and $n_{\rm add,gate}=0.0915$ quanta. The operator $L_{\rm out}$ describes emission into the external optical port. After adiabatic elimination of the resonant optical cavity, it takes the effective form
\begin{equation}
    L_{\rm out} = \sqrt{\eta_o\Gamma_{\rm om}}\,b,
    \quad
    \Gamma_{\rm om}=\frac{4|G_{\rm om}|^2}{\kappa},
    \quad
    \eta_o=\frac{\kappa_{\rm ext}}{\kappa}.
\end{equation}
The detected photon flux is therefore $\langle L_{\rm out}^{\dagger}L_{\rm out}\rangle =\eta_o\Gamma_{\rm om}\langle b^\dagger b\rangle$.

Uniform ZZ coupling preserves permutation symmetry but shifts the successive collective emission transitions, see Figs.~\ref{fig:robustness}(a) and \ref{fig:robustness}(c). For $H_{\rm ZZ}/\hbar=\zeta N_q(N_q-1)/2$, the transition from $j$ to $j-1$ qubit excitations acquires a shift $\zeta(j-1)$. Adding $-\zeta(k-1)N_q$ to $H/\hbar$ cancels this contribution for the initial $k=3$ transition. Subsequent transitions retain ZZ shifts $\zeta(j-k)$, so the integrated retrieval response depends on the full emission sequence. At $\zeta/g_{\rm sm}=-4$, first-transition retuning increases the efficiency from $43.2\%$ to $58.0\%$. At $\zeta/g_{\rm sm}=+4$, the corresponding efficiencies are $53.0\%$ and $44.2\%$. Thus, first-transition resonance alone does not determine the optimum frequency setting for collecting all three initial excitations. Nonuniform ZZ interactions introduce configuration-dependent energy shifts and break permutation symmetry, see Figs.~\ref{fig:robustness}(b) and \ref{fig:robustness}(d). Centering the pair couplings removes their mean while retaining the spatial inhomogeneity. Increasing the RMS disorder from zero to $\sigma_\zeta/g_{\rm sm}=4$ reduces the mean retrieval efficiency from $58.1\%$ to $35.4\%$ and increases the mean input-referred noise from $0.0915$ to $0.150$ quanta. The mean background count changes only slightly, from approximately $0.0532$ to $0.0531$ photons. The increased input-referred noise therefore arises mainly from the reduced signal-retrieval efficiency. Effects of ZZ interactions on loading into the mechanical mode are shown in Fig.~\ref{fig:ZZ_loading} (Appendix~\ref{app:robustness_retrieval}).

The pump sweep shows a tradeoff between optical extraction and the prescribed pump-induced qubit rates, see Figs.~\ref{fig:robustness}(e)--\ref{fig:robustness}(h). With intrinsic qubit rates alone, the sampled retrieval efficiency reaches approximately $63.8\%$ near $n_c=274$. Including relaxation, dephasing, and excitation shifts the sampled maximum to approximately $52.9\%$ near $n_c=150$. Pump-induced excitation has a particularly strong effect on the background. At $n_c=100$, adding excitation alone changes the efficiency from $58.1\%$ to $58.2\%$, while increasing $n_{\rm add,gate}$ from $0.0915$ to $0.200$ quanta. With all three added channels, the efficiency is $51.4\%$ and the added noise is $0.229$ quanta. The continuous calculations evaluate a different operating protocol, with a weak incident signal and an $N=6$ ensemble linearized about its ground state. We use
\begin{equation}
    \eta_{\rm CW}^{\rm peak} =\max_{\omega}|S_{o\leftarrow s}(\omega)|^2,
    \quad
    n_{\rm add,CW}=\frac{S_{\rm out}^{\rm noise}(\omega_\star)}{\eta_{\rm CW}(\omega_\star)},
\end{equation}
where $\omega_\star$ maximizes external conversion efficiency. The baseline response reaches approximately $37.6\%$ near $n_c=163$, while added relaxation and dephasing reduce the sampled maximum to $34.9\%$ near $n_c=127$. The included pump-dependent mechanical bath produces increasing spectral noise, reaching approximately $9.01$ quanta at $n_c=500$ for the baseline and $13.8$ quanta with both added qubit channels. These efficiencies measure different quantities. The continuous efficiency $\eta_{\rm CW}$ is the steady-state conversion efficiency of Fig.~\ref{fig:linear}, evaluated at the larger coupling of Fig.~\ref{fig:robustness}. The gated efficiency $\eta_{\rm gate}$ is instead the fraction of excitations stored in the ensemble that are emitted into the external port within $T_g$, and it does not involve converting an incoming signal. The two calculations also differ in initial state and in mechanical heating, so their values should not be compared directly.

\section{Discussion \& Conclusion}
We have investigated a collective superconducting interface for mechanically mediated microwave-to-optical quantum transduction. In the weak-excitation regime, the full input--output response connects collective ensemble--mechanical coupling to external conversion efficiency, added noise, and spectral bandwidth. The resulting operating points depend on ensemble size, optical drive, extraction losses, and mechanical noise. Beyond this regime, the finite Dicke ladder provides an excitation-dependent interaction whose selected matrix elements approach linear scaling with ensemble size near half excitation. The nonlinear calculations examine how mechanical anharmonicity enables selective transfer and how local decoherence and frequency disorder constrain that operation. The robustness calculations further distinguish common interaction shifts, which can be compensated at the selected transition, from pair-dependent interactions that distort the collective dynamics. The optical-retrieval calculations show that pump-induced qubit noise limits the useful pump range, with qubit excitation mainly increasing the added noise.

One route to a realizable device is the integration of a tunable superconducting-qubit ensemble with a common high-frequency mechanical mode embedded in an optomechanical crystal. Such an architecture would require approximately uniform qubit–mechanical coupling, controlled frequency disorder, and sufficient mechanical anharmonicity to resolve the desired transfer channel. Recent advances in superconducting electromechanics and mechanically induced quantum nonlinearities provide relevant experimental building blocks~\cite{Bozkurt2023LongLivedPhonons,Yang2024MechanicalQubit}. Further improvements may be obtained through optimized pulse shaping, dynamically controlled superconducting ensemble–mechanical interactions, and selective addressing of individual Dicke transitions. For the half-excited encoding of Eq.\ \eqref{eq:encoding}, further work is needed to connect coherent encoding, selective extraction, and the temporal mode of the propagating optical field within one complete channel description~\cite{Zeuthen2020FiguresMerit,Kiilerich2019QuantumPulses}.

Collective Dicke enhancement can speed up electromechanical transfer, but only if $g_{\mathrm{sm}}T_1$ is large and the mechanical anharmonicity exceeds the collective coupling. In contrast to approaches that only raise cooperativity, the ensemble lets the coupling be tuned through the excitation number. Combined with engineered mechanical nonlinearities, collective superconducting states could provide reconfigurable mechanically mediated links between superconducting quantum processors.

\begin{acknowledgements}
We acknowledge helpful discussions with Murat Can Sarihan, Amina El Ghazi, Jugal Talukdar, Alessandro Reineri, Cody Fan, and Silvia Zorzetti. This work was supported by the Defense Advanced Research Projects Agency (DARPA), Quantum Science Center (a National Quantum Information Science Center of the U.S. Department of Energy), Gordon and Betty Moore Foundation (Grant Numbers GBMF8048 and GBMF12976), and the John Simon Guggenheim Memorial Foundation (Guggenheim Fellowship).
\end{acknowledgements}

\onecolumngrid
\appendix
\section{Dicke States and Optomechanical Transducers}
\label{sec:background}
\begin{figure}[h]
    \centering
    \includegraphics[width=\linewidth]{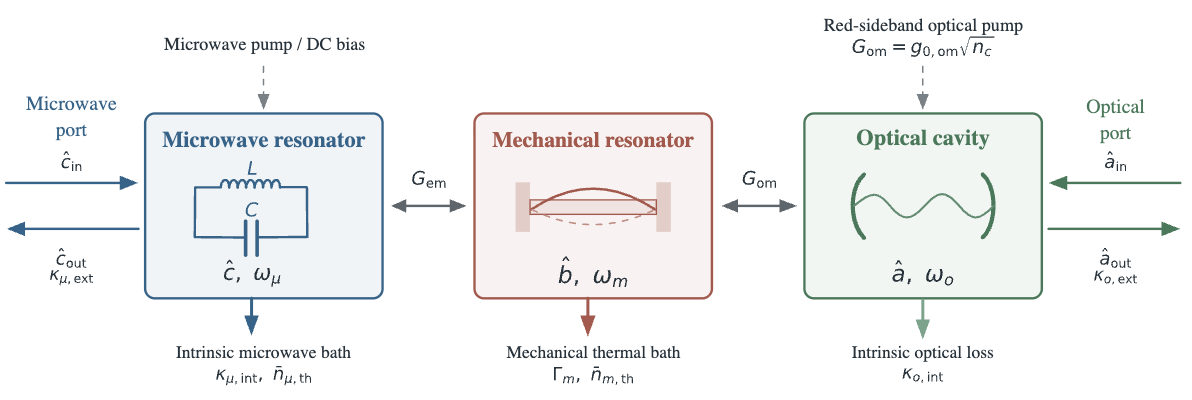}
    \caption{Conventional mechanically mediated microwave-to-optical transduction. A microwave LC mode ($\hat c$) couples to an optical mode ($\hat a$) through the mechanical mode ($\hat b$), with exchange rates $G_{\mathrm{em}}$ and $G_{\mathrm{om}}$. The diagram distinguishes external ports from intrinsic losses and includes mechanical thermalization and red-sideband optical driving. Optical absorption can introduce additional mechanical heating.}
    \label{fig:generic_transducer}
\end{figure}

\subsection{Optomechanical Transducer}
A conventional optomechanical transducer uses a mechanical resonator to mediate coherent exchange between a microwave resonator and an optical cavity, as illustrated in Fig.~\ref{fig:generic_transducer}. The microwave--mechanical interaction can arise from piezoelectric coupling, voltage-biased capacitive coupling, or a parametrically driven electromechanical interaction. For example, electrostatic actuation allows the coupling to be tuned through a DC voltage without requiring piezoelectric materials~\cite{Zhao2025QuantumEnabledNanomechanics}. The optical interaction originates from the displacement dependence of the cavity resonance~\cite{Aspelmeyer2014CavityOptomechanics}. After linearizing the optical field about a coherent pump and applying the rotating-wave approximation on the red mechanical sideband, the near-resonant Hamiltonian takes the form
\begin{equation}
\begin{aligned}
\frac{\hat H_{\mathrm{lin}}}{\hbar} ={}&\delta_\mu\hat c^\dagger\hat c +\delta_m\hat b^\dagger\hat b +\delta_o\hat a^\dagger\hat a +G_{\mathrm{em}} (\hat c^\dagger\hat b+\hat c\hat b^\dagger)+G_{\mathrm{om}} (\hat a^\dagger\hat b+\hat a\hat b^\dagger),
\end{aligned}
\label{eq:review_linear_hamiltonian}
\end{equation}
where \(\hat c\), \(\hat b\), and \(\hat a\) annihilate microwave, mechanical, and optical signal excitations, respectively, and \(\delta_j\) are residual detunings from the frequency-matching condition. The pump-enhanced optical coupling is \(G_{\mathrm{om}}=g_{0,\mathrm{om}}\sqrt{n_c}\), with \(g_{0,\mathrm{om}}\) the single-photon coupling and \(n_c\) the intracavity pump occupation. Both exchange terms conserve signal excitation number within this approximation. An isolated resonant mode pair with constant coupling \(G\) undergoes a complete single-excitation swap in \(t_{\mathrm{swap}}=\pi/(2G)\). Conversion may instead be operated continuously, with both interactions active, or sequentially with mechanical loading preceding optical readout~\cite{Mirhosseini2020QubitOptical}.

For a passive, resonant, stationary converter, the external efficiency is
\begin{equation}
\eta(0)=\eta_\mu\eta_o \frac{4C_{\mathrm{em}}C_{\mathrm{om}}} {(1+C_{\mathrm{em}}+C_{\mathrm{om}})^2},
\label{eq:review_conversion_efficiency}
\end{equation}
where \(\eta_j=\kappa_{j,\mathrm{ext}}/\kappa_j\), \(C_{\mathrm{em}}=4G_{\mathrm{em}}^2/(\kappa_\mu\Gamma_m)\), and \(C_{\mathrm{om}}=4G_{\mathrm{om}}^2/(\kappa_o\Gamma_m)\). Here \(\kappa_j=\kappa_{j,\mathrm{ext}}+\kappa_{j,\mathrm{int}}\) are total cavity energy-decay rates and \(\Gamma_m\) is the mechanical energy-decay rate. $C_{om}$ and $C_{em}$ represent the optomechanical and electromechanical cooperativities. Efficient conversion therefore requires good extraction and approximately matched, large cooperativities. Mechanical thermal excitation, intrinsic losses, and pump-induced heating limit the conversion performance~\cite{Zeuthen2020FiguresMerit,Zhao2025QuantumEnabledNanomechanics}. When both cavity modes can be adiabatically eliminated, the conversion bandwidth is approximately \(B\simeq\Gamma_m(1+C_{\mathrm{em}}+C_{\mathrm{om}})/(2\pi)\), which fails when the system is pushed into a strong-coupling regime resulting in hybridized peaks. Mechanical dark-mode protocols can reduce sensitivity to mechanical dissipation, but their performance depends on the pulse sequence and available interaction strengths~\cite{wang2012darkmodes,dong2012darkmode}.

\subsection{Dicke states and collective superconducting ensemble--mechanical exchange}
An ensemble of \(N\) two-level systems supports permutation-symmetric Dicke states $k$ excitations,
\begin{equation}
\ket{D_k^{N}}=
\binom{N}{k}^{-1/2}
\sum_{\pi\in\mathcal P_{N,k}}
\pi\!\left(\ket{0}^{\otimes(N-k)}\otimes\ket{1}^{\otimes k}\right),
\label{eq:review_dicke_definition}
\end{equation}
where \(\mathcal P_{N,k}\) contains the distinct permutations. Taking \(\ket{1}\) as the excited state gives \(\hat S_z=\tfrac12\sum_j\hat\sigma_z^{(j)}\), \(J=N/2\), and \(m=k-N/2\). Symmetry means invariance under qubit permutation, and does not require half excitation. The collective operators \(\hat S_\pm=\sum_j\hat\sigma_\pm^{(j)}\) act as~\cite{Dicke1954Coherence,Tavis1968NMolecule}
\begin{align}
\hat S_-\ket{D_k^{N}} &=\sqrt{k(N-k+1)}\ket{D_{k-1}^{N}},\quad \hat S_+\ket{D_k^{N}} =\sqrt{(k+1)(N-k)}\ket{D_{k+1}^{N}}.
\end{align}

\noindent For identical coupling magnitudes and phases, replacing the microwave resonator by this ensemble gives
\begin{equation}
\begin{aligned}
\frac{\hat H_{\mathrm{3}}}{\hbar}
={}&\delta_s\hat S_z+ \delta_m\hat b^\dagger\hat b+ \delta_o\hat a^\dagger\hat a +g_{\mathrm{sm}} (\hat S_+\hat b+\hat S_-\hat b^\dagger) +G_{\mathrm{om}} (\hat a^\dagger\hat b+\hat a\hat b^\dagger).
\end{aligned}
\label{eq:review_collective_hamiltonian}
\end{equation}
The coefficient \(g_{\mathrm{sm}}\) remains the single-qubit coupling and the enhancement follows from the matrix element
\begin{equation}
\begin{aligned}
&\frac{1}{\hbar}\left|\bra{D_{k-1}^{N},n+1} \hat H_{\mathrm{sm}}\ket{D_k^{N},n}\right|=g_{\mathrm{sm}}\sqrt{k(N-k+1)}\sqrt{n+1}.
\end{aligned} \label{eq:review_collective_matrix_element}
\end{equation}
For initially empty mechanics, \(G_{\mathrm{sm}}(k)=g_{\mathrm{sm}}\sqrt{k(N-k+1)}\) is \(g_{\mathrm{sm}}\sqrt N\) at \(k=1\) and approaches \(g_{\mathrm{sm}}N/2\) near half excitation. The normalized bright operator \(\hat B_-=\hat S_-/\sqrt N\) is approximately bosonic only near the collective ground state, where the occupied excitation numbers satisfy \(k\ll N\), also known as the Holstein--Primakoff approximation. At higher excitation, the full set of superconducting operators must be retained.

\section{Direct collective electromechanical interface}
\label{app:direct_collective_interface}

We consider $N$ electrically distinct superconducting qubits coupled to one mechanical normal mode that also has optomechanical participation. The qubits need not share a microwave resonator during transduction. Each qubit instead terminates in a separate coupling electrode sampling the same mechanical displacement or strain field. A candidate electrostatic layout comprises an extended silicon phononic structure with several independently wired transmon electrodes in the mechanically active region and an optical cavity in a separate portion of the same normal mode. The transmon islands remain electrically distinct: connecting every electrode to one superconducting island would not produce $N$ independent qubits. Electrode symmetry, DC-bias adjustment, and qubit frequency tuning are candidate controls of the coupling magnitudes and resonances.  All frequencies and Hamiltonian coefficients below are angular frequencies unless explicitly divided by $2\pi$. Let $\mathbf C(x)$ be the reduced capacitance matrix of the independent superconducting nodes and $\mathbf q_g(x)$ the signed bias-induced offset-charge vector. With the bias electrodes treated as fixed-voltage terminals, the charging Hamiltonian is
\begin{equation}
H_C(x) = \frac12 [\mathbf Q-\mathbf q_g(x)]^T \mathbf C^{-1}(x) [\mathbf Q-\mathbf q_g(x)].
\end{equation}
Scalar bias-force terms are included in the mechanical equilibrium potential. Define $\widetilde{\mathbf Q}=\mathbf Q-\mathbf q_g(0)$ and $\mathbf C_0=\mathbf C(0)$. Using $\partial_x\mathbf C^{-1} =-\mathbf C^{-1}\mathbf C'\mathbf C^{-1}$, the first-order expansion is
\begin{align}
H_C(x)&=H_C(0)+x\hat A_m+O(x^2),\\
\hat A_m
&=
-\mathbf q_g'{}^T\mathbf C_0^{-1}
\widetilde{\mathbf Q}
-\frac12
\widetilde{\mathbf Q}^{T}
\mathbf C_0^{-1}\mathbf C_0'
\mathbf C_0^{-1}\widetilde{\mathbf Q}.
\end{align}
For weakly hybridized local qubits, the transverse electromechanical matrix element is
\begin{equation}
g_{im}
=
\frac{x_{\mathrm{zpf}}}{\hbar}
\langle e_i|\hat A_m|G\rangle,
\qquad
|G\rangle=|g_1\cdots g_N\rangle.
\end{equation}
For an approximately independent transmon,
\begin{equation}
H_{q,i}(x)
=
4E_{C,i}(x)
[\hat n_i^{\mathrm{ch}}-n_{g,i}(x)]^2
-E_{J,i}\cos\hat\varphi_i,
\end{equation}
so that, with
$\hat N_i=\hat n_i^{\mathrm{ch}}-n_{g,i}(0)$,
\begin{equation}
\left.\partial_x H_{q,i}\right|_0
=
4E'_{C,i}\hat N_i^2
-
8E_{C,i}n'_{g,i}\hat N_i.
\end{equation}
At a charge-symmetric operating point, the leading ground-to-excited-state matrix element is supplied by the charge-odd second term. With $E_{C,i}=e^2/(2C_{\Sigma,i})$ and $n'_{g,i}=-V_{\mathrm{dc},i}C'_{im}/(2e)$ for the chosen offset-charge convention, this yields Eq.\ \eqref{eq:direct_qubit_mechanical_coupling}. The off-diagonal inverse-capacitance elements already imply electrical interactions. In a local-transmon approximation their transverse matrix elements are
\begin{equation}
 J_{ij}\simeq \frac{(2e)^2}{\hbar}(\mathbf C_0^{-1})_{ij}
                    n_{eg,i}n_{ge,j},\qquad i\ne j.
\end{equation}
The quadratic electrical circuit does not by itself determine a qubit cross-Kerr. Conditional shifts also require the Josephson anharmonicity and, generally, multilevel diagonalization. A useful unambiguous extraction is
\begin{equation}
 \zeta_{ij}=\frac{E_{11}-E_{10}-E_{01}+E_{00}}{\hbar}.
\end{equation}

\section{Dispersive elimination of the auxiliary--mechanical interaction}
\label{app:aux_disp}
We begin from
\begin{equation}
\frac{\hat H_{am}}{\hbar}=\frac{(\hat H_{0,am}+\hat H_{\rm JC})}{\hbar} = \omega_m\hat b^\dagger\hat b + \frac{\omega_{\rm aux}}{2}\hat\sigma_z + g_{\rm aux} \left( \hat\sigma_+\hat b+ \hat\sigma_-\hat b^\dagger \right)
\end{equation}
To facilitate the transformation, we define the combined operators as $\hat X=\hat\sigma_+\hat b$, $\hat Y=\hat\sigma_-\hat b^\dagger$ and $\hat V=\lambda_{\rm aux}(\hat X-\hat Y)$ with $\lambda_{aux} = \frac{g_{aux}}{\Delta_{aux}}$ and $\Delta_{aux} = \omega_{aux}-\omega_m$. The BCH expansion is
\begin{equation}
e^{\hat V}\hat H_{am}e^{-\hat V} = \hat H_{am} + [\hat V,\hat H_{am}] +
\frac{1}{2} [\hat V,[\hat V,\hat H_{am}]] +\cdots.
\end{equation}
Using the relations $[\hat H_{0,am},\hat X] = \hbar\Delta_{\rm aux}\hat X$ and $[\hat H_{0,am},\hat Y] = -\hbar\Delta_{\rm aux}\hat Y$, we obtain
\begin{equation}
[\hat V,\hat H_{0,am}]
=
-\hat H_{\rm JC},
\end{equation}
so the direct auxiliary--mechanical exchange cancels to first order. The second-order contribution is $\hat H_{am}^{(2)} = \hat H_{0,am} + \frac{1}{2} [\hat V,\hat H_{\rm JC}]. $ The commutator is $[\hat X,\hat Y] = \frac{1}{2} + \left( \hat n+\frac{1}{2} \right)\hat\sigma_z$, and we obtain the full auxiliary-mechanical Hamiltonian
\begin{equation}
\frac{\hat H_{am}^{(2)}}{\hbar}
=
\omega_m\hat n
+
\frac{\omega_{\rm aux}}{2}\hat\sigma_z
+
\frac{g_{\rm aux}^2}{\Delta_{\rm aux}}
\left(
\hat n+\frac{1}{2}
\right)\hat\sigma_z
+{\rm const.}
\end{equation}
For the fourth-order diagonal correction, we introduce the conserved excitation number $\hat{\mathcal N} = \hat n+\frac{1+\hat\sigma_z}{2}.$ The exact Jaynes--Cummings eigenenergies in excitation manifold $q\ge1$ are
\begin{equation}
\frac{E_{q,\pm}}{\hbar}
=
\omega_m\left(q-\frac{1}{2}\right)
\pm
\frac{1}{2}
\sqrt{
\Delta_{\rm aux}^2+4g_{\rm aux}^2q
}.
\end{equation}
In the dispersive regime with $\Delta_{aux}>0$,
\begin{equation}
\frac{1}{2}
\sqrt{
\Delta_{\rm aux}^2+4g_{\rm aux}^2q
}
=
\frac{\Delta_{\rm aux}}{2}
+
\frac{g_{\rm aux}^2q}{\Delta_{\rm aux}}
-
\frac{g_{\rm aux}^4q^2}{\Delta_{\rm aux}^3}
+
\mathcal O\left(
\frac{g_{\rm aux}^6q^3}{\Delta_{\rm aux}^5}
\right).
\end{equation}
Applying the conserved excitation number operator $\mathcal{N}$ onto the auxiliary operators using  $\hat{\mathcal N}\hat\sigma_z = \hat n\hat\sigma_z+\frac{\hat\sigma_z+1}{2}$ and $\hat{\mathcal N}^2\hat\sigma_z = \hat n^2\hat\sigma_z + \hat n\hat\sigma_z + \hat n + \frac{\hat\sigma_z+1}{2},$ gives the dispersive auxiliary-mechanical Hamiltonian
\begin{align}
\frac{\hat H_{am}^{\rm disp}}{\hbar} ={}& \left( \omega_m-\frac{g_{\rm aux}^4}{\Delta_{\rm aux}^3} \right)\hat n + \frac{\omega_{\rm aux}}{2}\hat\sigma_z + \left( \frac{g_{\rm aux}^2}{\Delta_{\rm aux}} - \frac{g_{\rm aux}^4}{\Delta_{\rm aux}^3} \right) \left(\hat n+\frac{1}{2} \right)\hat\sigma_z - \frac{g_{\rm aux}^4}{\Delta_{\rm aux}^3} \hat n^2\hat\sigma_z .
\end{align}
This reproduces the definitions of $\chi_{\rm aux}$ and $K_{\rm m}$ used in the main text. Because the Schrieffer--Wolff transformation acts on $\hat b$ and $\hat b^\dagger$, it also transforms the Dicke--mechanical and optomechanical couplings. Using $[\hat V,\hat b] = \lambda_{\rm aux}\hat\sigma_-$ and $[\hat V,\hat b^\dagger] = \lambda_{\rm aux}\hat\sigma_+$, we find
\begin{align}
e^{\hat V}\hat H_{sm}e^{-\hat V}
={}&
\hat H_{sm}
+
\hbar g_{sm}\lambda_{\rm aux}
\left(
\hat S_+\hat\sigma_-
+
\hat S_-\hat\sigma_+
\right)
+
\frac{\hbar g_{sm}\lambda_{\rm aux}^2}{2}
\hat\sigma_z
\left(
\hat S_+\hat b+\hat S_-\hat b^\dagger
\right)
+
\mathcal O(\lambda_{\rm aux}^3),
\\
e^{\hat V}\hat H_{om}e^{-\hat V}
={}&
\hat H_{om}
+
\hbar G_{om}\lambda_{\rm aux}
\left(
\hat a^\dagger\hat\sigma_-
+
\hat a\hat\sigma_+
\right)
+
\frac{\hbar G_{om}\lambda_{\rm aux}^2}{2}
\hat\sigma_z
\left(
\hat a^\dagger\hat b+\hat a\hat b^\dagger
\right)
+
\mathcal O(\lambda_{\rm aux}^3).
\end{align}
The leading induced interaction scales as $G_{s{\rm aux}}^{\rm ind} \sim \frac{g_{sm}g_{\rm aux}}{\Delta_{\rm aux}}$ and $G_{o{\rm aux}}^{\rm ind} \sim \frac{G_{om}g_{\rm aux}}{\Delta_{\rm aux}}$. The same transformation applies to the dissipative operators: $\widetilde L_\mu=e^{\hat V}L_\mu e^{-\hat V}$. A microscopic effective master equation therefore requires the Hamiltonian and dissipators to be transformed consistently. Alternatively, the auxiliary interaction and its local dissipators can be retained explicitly in the bare representation.

\begin{figure}[h]
    \centering
    \includegraphics[width=\linewidth]{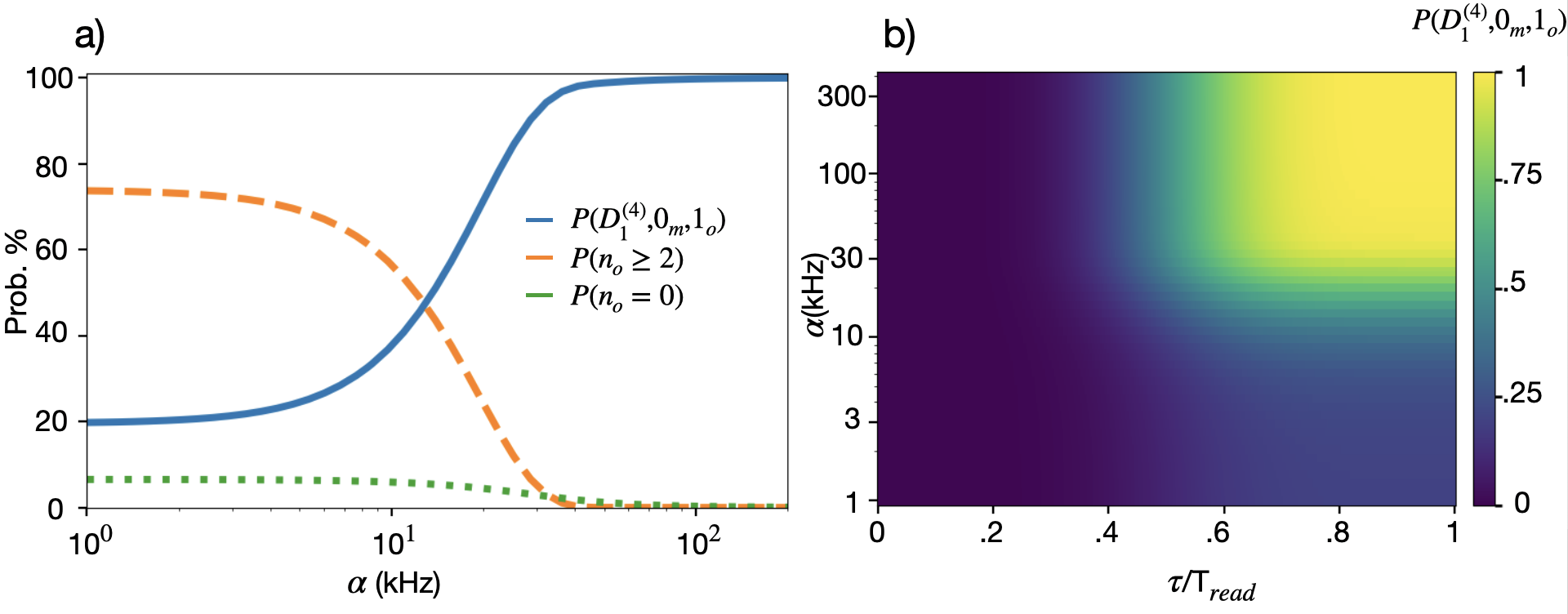}
    \caption{Dependence of selective intracavity transfer on the prescribed mechanical anharmonicity. The $N=4$ ensemble is initially prepared in $\lvert D_2^{4},0_m,0_o\rangle$ and undergoes lossless mechanical loading followed by optical readout. \textbf{a)} Final joint-state probability $P_{\mathrm{tar}}=P(D_1^{4},0_m,1_o)$ (blue solid), optical multiphoton probability $P(n_o\geq2)$ (orange dashed), and optical vacuum probability $P(n_o=0)$ (green dotted), expressed as percentages.
    \textbf{b)} Absolute target-state probability during optical readout, with $\tau$ measured from the beginning of the readout pulse. The nonoverlapping $\sin^2$ coupling pulses have fixed amplitudes and durations throughout the anharmonicity sweep, with $g_{\mathrm{sm}}/(2\pi)=3.81~\mathrm{kHz}$ and $G_{\mathrm{om}}/(2\pi)=3.43~\mathrm{MHz}$. }
    \label{fig:alpha}
\end{figure} 
\vspace{1cm}
Figure~\ref{fig:alpha} connects mechanical transition selectivity to the optical outcome of the complete sequential protocol. The heatmap shows the absolute joint-state probability $P_{\mathrm{tar}}(\tau)=P(D_1^{4},0_m,1_o)$ during optical readout as a function of the prescribed mechanical anharmonicity, where $\tau$ is measured from the start of the optical pulse. For each anharmonicity, the full state obtained after mechanical loading is passed to readout without projection or renormalization. The ensemble--mechanical interaction is switched off while the Kerr term remains active. Both pulse envelopes, their peak amplitudes, and their durations are held fixed. At small anharmonicity, nonselective loading limits the population reaching the desired optical state. Increasing the separation of the mechanical transitions suppresses this loading error and produces a high-probability plateau at the end of readout. For $N=4$, $k=2$, and $g_{\mathrm{sm}}/(2\pi)=3.81~\mathrm{kHz}$, $\alpha_m/(2\pi)=100~\mathrm{kHz}$ gives a final optical target probability of $99.71\%$, compared with $19.58\%$ in the harmonic case. The corresponding final optical multiphoton probability is $3.85\times10^{-6}$. The heatmap therefore gives the anharmonicity required for selective intracavity transfer under fixed controls.

\section{Rotating frame and dressed resonance conditions}
\label{app:rotating_frame}

A common excitation frame may be defined by
\begin{equation}
\hat U_r(t)
=
\exp\left[
i\omega_r t
\left(
\hat S_z+
\hat b^\dagger\hat b+
\hat a^\dagger\hat a+
\frac{\hat\sigma_z}{2}
\right)
\right].
\end{equation}
The rotating-frame Hamiltonian is $\hat H_r = \hat U_r\hat H\hat U_r^\dagger + i\hbar\dot{\hat U}_r\hat U_r^\dagger .$ The relevant transformations are
\begin{align}
\hat U_r\hat S_\pm\hat U_r^\dagger
&=
e^{\pm i\omega_r t}\hat S_\pm,
\quad
\hat U_r\hat b\hat U_r^\dagger
=
e^{-i\omega_r t}\hat b,
\quad
\hat U_r\hat a\hat U_r^\dagger
=
e^{-i\omega_r t}\hat a,
\quad
\hat U_r\hat\sigma_\pm\hat U_r^\dagger
=
e^{\pm i\omega_r t}\hat\sigma_\pm.
\end{align}
Hence the exchange terms remain time independent, and
\begin{align}
\frac{\hat H_r}{\hbar}
={}&
\delta_s\hat S_z
+
\delta_m\hat b^\dagger\hat b
+
\delta_o\hat a^\dagger\hat a
+
\frac{\delta_{\rm aux}}{2}\hat\sigma_z
+
g_{sm}
\left(
\hat S_+\hat b+\hat S_-\hat b^\dagger
\right)
+
G_{om}
\left(
\hat a^\dagger\hat b+\hat a\hat b^\dagger
\right)
+
g_{\rm aux}
\left(
\hat\sigma_+\hat b+\hat\sigma_-\hat b^\dagger
\right),
\end{align}
with $\delta_s=\omega_s-\omega_r$, $\delta_m=\omega_m-\omega_r$, $\delta_o=-\Delta_o-\omega_r$, and $\delta_{\rm aux}=\omega_{\rm aux}-\omega_r$ the detunings. Expressed in laboratory-frequency notation, the ground-auxiliary mechanical Hamiltonian and energy eigenstates are
\begin{equation}
\frac{\hat H_m^{(g)}}{\hbar}
=
\left(
\omega_m+K_{\rm m}-\chi_{\rm aux}
\right)\hat n
-
K_{\rm m}\hat n^2,
\quad
\frac{E_n^{(g)}}{\hbar}
=
\left(
\omega_m+K_{\rm m}-\chi_{\rm aux}
\right)n
-
K_{\rm m}n^2
+{\rm const.},
\end{equation}
up to a constant  and $\omega_{n,n+1}^{(g)} = \omega_m-\chi_{\rm aux}-2n K_{\rm m}$. In particular,
\begin{align}
\omega_{0,1}^{(g)}
&= \omega_m-\chi_{\rm aux}, \quad \omega_{1,2}^{(g)}
= \omega_m-\chi_{\rm aux}-2K_{\rm m}, \quad
\omega_{2,3}^{(g)} = \omega_m-\chi_{\rm aux}-4K_{\rm m}.
\end{align}
The first Dicke transition should therefore be tuned approximately to $\omega_s\simeq -\Delta_o \simeq \omega_{0,1}^{(g)} =  \omega_m-\chi_{\rm aux}$ when the auxiliary qubit remains in its ground state.


\section{Full input--output state-space representation}
\label{app:io_matrices}

\subsection{State-space scattering representation}
\label{app:state_space_scattering}
For either architecture the linearized equations may be written
\begin{equation}
\dot{\mathbf v}(t)
=
\mathbf A\mathbf v(t)
+
\mathbf B\mathbf v_{\mathrm{in}}(t),
\qquad
\mathbf v_{\mathrm{out}}(t)
=
\mathbf C\mathbf v(t)
+
\mathbf D\mathbf v_{\mathrm{in}}(t).
\end{equation}
With the Fourier convention used above,
\begin{equation}
\mathbf S(\omega)
=
\mathbf C
\left(
-i\omega\mathbf I-\mathbf A
\right)^{-1}
\mathbf B
+
\mathbf D.
\label{eq:general_scattering_matrix}
\end{equation}

Let $\Gamma_m=\Gamma_h+\Gamma_f$ describe the two intrinsic mechanical baths. For a given device, write $\Gamma=\Gamma_m+\Gamma_A$, taking $\Gamma_A=0$ for the three-mode case. The Langevin equations of motion are
\begin{align}
 \dot s={}&-\left(\frac{\gamma_s}{2}+i\delta_s\right)s-iG_{\rm sm}b +\sqrt{\gamma_{s,\rm ext}}s_{\rm in,ext}
 +\sqrt{\gamma_{s,\rm int}}s_{\rm in,int},\\
 \dot b={}&-\left(\frac{\Gamma}{2}+i\delta_m\right)b
 -iG_{\rm sm}s-iG_{\rm om}a +\sqrt{\Gamma_h}b_h+\sqrt{\Gamma_f}b_f+\sqrt{\Gamma_A}b_A,\\
 \dot a={}&-\left(\frac{\kappa}{2}+i\delta_o\right)a-iG_{\rm om}b +\sqrt{\kappa_{\rm ext}}a_{\rm in,ext}
 +\sqrt{\kappa_{\rm int}}a_{\rm in,int}
 \label{eq:io_langevin}
\end{align}

where $s = \frac{S_-}{\sqrt{N}}$, $b$ and $a$ are the respective superconducting ensemble, mechanical and optical operators.

\subsection{Three-mode system}
For the red-sideband three-mode system, define the state vector
$\mathbf v_-^{(3)}
=
\begin{pmatrix}
\hat s & \hat b & \hat a
\end{pmatrix}^{T}$, and the input vector
$\mathbf v_{\mathrm{in},-}^{(3)}
= \begin{pmatrix}
\hat s_{\mathrm{ext}} &
\hat s_{\mathrm{int}} &
\hat b_h &
\hat b_f &
\hat a_{\mathrm{ext}} &
\hat a_{\mathrm{int}}
\end{pmatrix}^{T}.$
The Langevin equations may be written
\begin{equation}
\dot{\mathbf v}_-^{(3)}
=
\mathbf A_-^{(3)}\mathbf v_-^{(3)}
+
\mathbf B_-^{(3)}\mathbf v_{\mathrm{in},-}^{(3)},
\end{equation}
with
\begin{equation}
\mathbf A_-^{(3)}
=
\begin{pmatrix}
-\left(i\delta_s+\gamma_s/2\right)
&
-iG_{sm}
&
0
\\[1mm]
-iG_{sm}
&
-\left(i\delta_m+\Gamma_m/2\right)
&
-iG_{om}
\\[1mm]
0
&
-iG_{om}
&
-\left(i\delta_o+\kappa/2\right)
\end{pmatrix}, 
\quad
\mathbf B_-^{(3)}
=
\begin{pmatrix}
\sqrt{\gamma_{s,\mathrm{ext}}}
&
\sqrt{\gamma_{s,\mathrm{int}}}
&
0
&
0
&
0
&
0
\\[1mm]
0
&
0
&
\sqrt{\Gamma_h}
&
\sqrt{\Gamma_f}
&
0
&
0
\\[1mm]
0
&
0
&
0
&
0
&
\sqrt{\kappa_{\mathrm{ext}}}
&
\sqrt{\kappa_{\mathrm{int}}}
\end{pmatrix}.
\label{eq:Aminus3_appendix}
\end{equation}
Here $\gamma_s=\gamma_{s,\mathrm{ext}}+\gamma_{s,\mathrm{int}}$ and $\kappa=\kappa_{\mathrm{ext}}+\kappa_{\mathrm{int}}$. For a passive, number-conserving red-sideband Hamiltonian, the creation sector is the complex-conjugate block. Using the boundary conditions $\hat s_{\mathrm{out}} = \hat s_{\mathrm{in}} - \sqrt{\gamma_{s,\mathrm{ext}}}\hat s$ and  $\hat a_{\mathrm{out}} = \hat a_{\mathrm{in}} - \sqrt{\kappa_{\mathrm{ext}}}\hat a$ and applying the same boundary condition to every port, the annihilation-sector outputs are
\begin{equation}
\mathbf v_{\mathrm{out},-}^{(3)}
=
\mathbf v_{\mathrm{in},-}^{(3)}
-
\mathbf B_{-}^{(3)T}\mathbf v_{-}^{(3)},
\end{equation}
whose $s_{\mathrm{ext}}$ and $a_{\mathrm{ext}}$ entries reproduce the boundary conditions above.
Different sign conventions for the input--output boundary condition alter only overall scattering phases and not the conversion efficiency.

\subsection{Four-mode system}

In the weak-excitation limit for the auxiliary qubit,
$\hat\sigma_z\simeq-1$, define $\mathbf v_-^{(4)} =
\begin{pmatrix}
\hat s & \hat b & \hat a & \hat\sigma_-
\end{pmatrix}^{T}$.
The annihilation-sector drift matrix is
\begin{equation}
\mathbf A_-^{(4)}
=
\begin{pmatrix}
-\left(i\delta_s+\gamma_s/2\right)
&
-iG_{sm}
&
0
&
0
\\[1mm]
-iG_{sm}
&
-\left(i\delta_m+\Gamma_m/2\right)
&
-iG_{om}
&
-ig_{\mathrm{aux}}
\\[1mm]
0
&
-iG_{om}
&
-\left(i\delta_o+\kappa/2\right)
&
0
\\[1mm]
0
&
-ig_{\mathrm{aux}}
&
0
&
-\left(i\delta_{\mathrm{aux}}+\Gamma_{2,\mathrm{aux}}\right)
\end{pmatrix}.
\label{eq:nadd_scattering_appendix}
\end{equation}

Auxiliary pure dephasing contributes to the coherence susceptibility, but its noise cannot be specified merely by assigning an independent bosonic thermal occupation. The noise calculation must use a consistent auxiliary reservoir model. Relaxation may be represented by a bath associated with $\gamma_{1,\mathrm{aux}}$, whereas pure dephasing enters the coherence rate $\Gamma_{2,\mathrm{aux}}=\frac{\gamma_{1,\mathrm{aux}}}{2}+\gamma_{\phi,\mathrm{aux}}$.


\section{Conversion efficiency and bandwidth}
\label{app:efficiency_bandwidth}

\subsection{Three-mode transfer function}

Using the Fourier convention $\hat o(t)=\int d\omega\,\hat o(\omega)e^{-i\omega t}/2\pi$, we define the susceptibilities $\chi_i^{-1}(\omega) = -i(\omega-\delta_i) + \frac{\kappa_i}{2}$:
\begin{align}
\chi_s^{-1}(\omega)
&=
\frac{\gamma_s}{2}
-i(\omega-\delta_s),
\quad
\chi_m^{-1}(\omega)
=
\frac{\Gamma_m}{2}
-i(\omega-\delta_m),
\quad
\chi_a^{-1}(\omega)
=
\frac{\kappa}{2}
-i(\omega-\delta_o).
\end{align}
All frequencies are measured in the common rotating frame of Appendix~\ref{app:rotating_frame}, so resonance corresponds to $\omega=0$ with $\delta_s=\delta_m=\delta_o=0$. The frequency-domain equations for a signal incident only through the external ensemble port are
\begin{align}
\chi_s^{-1} s
+iG_{sm}b
&=
\sqrt{\gamma_{s,\mathrm{ext}}}\,s_{\mathrm{in}},
\\
\chi_m^{-1} b
+iG_{\mathrm{sm}}s
+iG_{om}a
&=0,
\\
\chi_a^{-1} a
+iG_{om}b
&=0.
\end{align}
Eliminating $s$ and $b$ gives
\begin{equation}
a(\omega)
=
-\frac{
G_{sm}G_{om}
\sqrt{\gamma_{s,\mathrm{ext}}}
}{
D_3(\omega)
}
s_{\mathrm{in}}(\omega),
\end{equation}
where $D_3 = \chi_s^{-1}\chi_m^{-1}\chi_a^{-1} + G_{sm}^2\chi_a^{-1} + G_{om}^2\chi_s^{-1}$. With $a_{\mathrm{out}}=a_{\mathrm{in}}- \sqrt{\kappa_{\mathrm{ext}}}a$, the forward conversion amplitude is therefore Eq.\ \eqref{eq:Xi3_main}, up to an overall phase. The corresponding efficiency is
\begin{equation}
\eta_3(\omega)
=
\frac{
\gamma_{s,\mathrm{ext}}\kappa_{\mathrm{ext}}
G_{sm}^2G_{om}^2
}{
|D_3(\omega)|^2
}.
\label{eq:eta3_full_appendix}
\end{equation}

On exact resonance the susceptibilities reduce to $\chi_s^{-1}=\frac{\gamma_s}{2}$, $\chi_m^{-1}=\frac{\Gamma_m}{2}$ and $\chi_a^{-1}=\frac{\kappa}{2}$. Therefore the denominator $D_3(\omega)$ becomes $ D_3(0) = \frac{1}{8} \left(\gamma_s\Gamma_m\kappa + 4\kappa G_{sm}^2 + 4\gamma_sG_{om}^2 \right)$. Introducing the cooperativity terms $C_{sm} = \frac{4G_{sm}^2}{\gamma_s\Gamma_m}$ and $C_{om} = \frac{4G_{om}^2}{\kappa\Gamma_m}$ with the extraction efficiencies $\eta_s = \frac{\gamma_{s,ext}}{\gamma_s}$ and $\eta_o = \frac{\kappa_{ext}}{\kappa}$ gives the resonant efficiency parameter:
\begin{equation}
\eta_3(0)
=
\eta_s\eta_o
\frac{4C_{sm}C_{om}}
{\left(1+C_{sm}+C_{om}\right)^2}.
\end{equation}

\subsection{Three-mode bandwidth}

When the superconducting ensemble and optical modes respond rapidly enough to be eliminated over the conversion band, they generate the induced mechanical damping rates
\begin{equation}
\Gamma_{sm}= \frac{4G_{sm}^2}{\gamma_s}, \qquad \Gamma_{om} = \frac{4G_{om}^2}{\kappa}.
\end{equation}
The effective mechanical linewidth is $\Gamma_{\mathrm{tot}}^{(3)} = \Gamma_m+\Gamma_{sm}+\Gamma_{om}$, which gives the approximate bandwidth
\begin{equation}
B_3
\simeq
\frac{\Gamma_{\mathrm{tot}}^{(3)}}{2\pi}.
\end{equation}
When $G_{sm}$ is increased by the Dicke enhancement toward the
strong-coupling regime, this single-pole approximation need not remain valid.
The bandwidth should then be reported from the full width at half maximum of
$\eta_3(\omega)$ calculated from Eq.\ \eqref{eq:eta3_full_appendix}.

\subsection{Four-mode extension}

For the auxiliary mode define $\chi_q^{-1}(\omega) = \Gamma_{2,\mathrm{aux}} -i(\omega-\delta_{\mathrm{aux}})$. The frequency-domain auxiliary equation gives
\begin{equation}
\sigma_-
=
-i g_{\mathrm{aux}}\chi_q b
+
\chi_q F_{\mathrm{aux}}.
\end{equation}
Substitution into the mechanical equation yields
\begin{equation}
\left[
\chi_m^{-1}
+
g_{\mathrm{aux}}^2\chi_q
\right]b
+
iG_{\mathrm{sm}}s
+
iG_{om}a
=
\text{noise},
\end{equation}
so the auxiliary self-energy is $\Sigma_{\mathrm{aux}}(\omega) = g_{\mathrm{aux}}^2\chi_q(\omega)$. The four-mode denominator is $D_4 = \chi_q^{-1}D_3 + g_{\mathrm{aux}}^2 \chi_s^{-1}\chi_a^{-1}$ accounting for the additional terms stemming from the auxiliary qubit. Near the mechanical resonance,
\begin{equation}
\Sigma_{\mathrm{aux}}(\delta_m)
=
\frac{
g_{\mathrm{aux}}^2
}{
\Gamma_{2,\mathrm{aux}}+i\Delta_{\mathrm{aux}}
},
\qquad
\Delta_{\mathrm{aux}}
=
\omega_{\mathrm{aux}}-\omega_m.
\end{equation}
Separating real and imaginary parts gives the auxiliary-induced damping  and the dispersive mechanical frequency shift
\begin{equation}
\Gamma_{\mathrm{aux}}
\simeq
\frac{
2g_{\mathrm{aux}}^2\Gamma_{2,\mathrm{aux}}
}{
\Delta_{\mathrm{aux}}^2+\Gamma_{2,\mathrm{aux}}^2
},\quad
\delta\omega_m^{(\mathrm{aux})}
\simeq
-\frac{
g_{\mathrm{aux}}^2\Delta_{\mathrm{aux}}
}{
\Delta_{\mathrm{aux}}^2+\Gamma_{2,\mathrm{aux}}^2
}.
\end{equation}
In the far-dispersive limit this reduces to $\delta\omega_m^{(\mathrm{aux})}\simeq-g_{\mathrm{aux}}^2/\Delta_{\mathrm{aux}}$. The approximate four-mode linewidth is then
\begin{equation}
\Gamma_{\mathrm{tot}}^{(4)}
\simeq
\Gamma_m+\Gamma_{sm}+\Gamma_{om}+\Gamma_{\mathrm{aux}},
\end{equation}
while the reported bandwidth should be obtained from the FWHM of the complete four-mode spectrum $\eta_4(\omega)=|\Xi_{as}^{(4)}(\omega)|^2$ whenever the auxiliary produces appreciable spectral structure.

\section{Open-system dissipation, thermal reservoirs, and input-referred noise}
\label{app:thermal_occupancy}
\label{app:added_noise}
\label{app:master_equation}
This appendix first defines the microscopic dissipative channels used in the master-equation calculations and then specifies the reservoir description used for the weak-excitation input--output analysis. For the latter, the collective superconducting ensemble is represented by $\hat s\simeq\hat S_-/\sqrt N$ with $G_{\mathrm{sm}}=g_{\mathrm{sm}}\sqrt N$.
The external microwave input and the optical reservoirs are assumed to be in vacuum apart from the incident signal being characterized, while the intrinsic superconducting reservoir and the mechanical reservoirs may have finite occupations. All reservoirs are taken to be independent and Markovian, with approximately flat noise spectra over the relevant conversion band.

\subsection{Master equation and collapse operators}
We consider independent energy relaxation and pure dephasing of each superconducting qubit, described by
\begin{equation}
\hat L_{1,j}=\sqrt{\gamma_{1,j}}\,\hat\sigma_-^{(j)},
\qquad
\hat L_{\phi,j}=\sqrt{\frac{\gamma_{\phi,j}}{2}}\,\hat\sigma_z^{(j)},
\end{equation}
with relaxation and pure-dephasing rates $\gamma_{1,j}=\frac{1}{T_{1,j}}, \gamma_{\phi,j} =\frac{1}{T_{2,j}}-\frac{1}{2T_{1,j}}.$ Mechanical dissipation and thermal excitation are modeled through
\begin{equation}
\hat L_{m,\downarrow} =\sqrt{\Gamma_m(\bar n_{\mathrm{th}}+1)}\,\hat b, \qquad \hat L_{m,\uparrow} =\sqrt{\Gamma_m\bar n_{\mathrm{th}}}\,\hat b^\dagger,
\end{equation}
where \(\Gamma_m\) denotes the intrinsic mechanical energy-decay rate and  $\bar n_{\mathrm{th}} = \left[ \exp\left(\frac{\hbar\omega_m}{k_BT_m}\right)-1 \right]^{-1}$ is the equilibrium mechanical bath occupation at temperature \(T_m\). Optical dissipation is separated into external extraction and intrinsic cavity loss, 
\begin{equation}
\hat L_{o,\mathrm{ext}} =\sqrt{\kappa_{\mathrm{ext}}}\,\hat a,
\qquad
\hat L_{o,\mathrm{int}}=\sqrt{\kappa_{\mathrm{int}}}\,\hat a,
\end{equation}
with total optical linewidth $\kappa=\kappa_{\mathrm{ext}}+\kappa_{\mathrm{int}}$. Finally, relaxation and dephasing of the auxiliary superconducting qubit are described by
\begin{equation}
\hat L_{\mathrm{aux},1} =\sqrt{\gamma_{1,\mathrm{aux}}}\, \hat\sigma_-^{(\mathrm{aux})},
\qquad
\hat L_{\mathrm{aux},\phi} =\sqrt{\frac{\gamma_{\phi,\mathrm{aux}}}{2}}\, \hat\sigma_z^{(\mathrm{aux})}.
\end{equation}
By retaining the auxiliary-mechanical exchange interaction explicitly, the bare auxiliary model captures the hybridization and associated dissipative dynamics without introducing an independently prescribed Kerr nonlinearity.

\subsection{Mechanical reservoirs and optical-absorption heating}

The mechanical mode interacts with a cold refrigerator bath, characterized by the energy-decay rate $\Gamma_f$ and occupation $n_f$, and an additional phonon bath generated by absorption of the optical pump, characterized by $\Gamma_h(n_c)$ and $n_h(n_c)$~\cite{Zhao2025QuantumEnabledNanomechanics}. Here $n_c$ is the intracavity optical pump occupation. The hot bath describes heating of the mechanical environment, rather than a thermal population of the optical cavity. The total intrinsic mechanical damping and thermal injection rate are
\begin{align}
\Gamma_m(n_c)
&=\Gamma_f+\Gamma_h(n_c),
\quad
Q_m(n_c)=\Gamma_f n_f+\Gamma_h(n_c)n_h(n_c)
\end{align}
Their ratio defines the occupation of the equivalent intrinsic mechanical bath, $\bar n_m(n_c)=\frac{Q_m(n_c)}{\Gamma_m(n_c)}$. This quantity is distinct from the mechanical-mode occupation in the operating converter, which also depends on exchange with the superconducting ensemble and optical subsystems. In the bare-frame master equation, the two mechanical baths give
\begin{equation}
\begin{aligned}
\mathcal L_m\rho ={}& \sum_{\nu=f,h}\Gamma_\nu(n_\nu+1) \mathcal D[\hat b]\rho+\sum_{\nu=f,h}\Gamma_\nu n_\nu \mathcal D[\hat b^\dagger]\rho,
\end{aligned}
\label{eq:two_bath_mechanical_dissipation}
\end{equation}
where $\mathcal D[\hat L]\rho= \hat L\rho\hat L^\dagger -\{\hat L^\dagger\hat L,\rho\}/2$. Equivalently, the combined collapse operators are
\begin{equation}
\hat L_{m,\downarrow}
=\sqrt{\Gamma_m+Q_m}\,\hat b,
\qquad
\hat L_{m,\uparrow}
=\sqrt{Q_m}\,\hat b^\dagger.
\label{eq:combined_mechanical_collapses}
\end{equation}
Thus, absorption heating modifies both the intrinsic damping and the thermal noise injection. Optical extraction is not included in $\Gamma_m$ when the optical cavity and its coupling are retained explicitly, since this would double count the same interaction. For the heating sensitivity calculations, we adopt
\begin{equation}
\frac{\Gamma_h(n_c)}{2\pi} =(1.1\,\mathrm{kHz})\,n_c^{0.61},
\qquad
n_h(n_c)=1.1\,n_c^{0.30},
\label{eq:hot_bath_power_laws}
\end{equation}
motivated by measurements in silicon optomechanical crystals~\cite{ren2020two}. These coefficients define a reference heating scenario, not a calibration of the proposed device. The damping power law was fitted over the range $100<n_c<1000$, with its use below this range extrapolated. The separate frequency-jitter contribution reported in that experiment is not included as an energy-decay channel.

\subsection{Steady-state occupations}
For the uncoupled ensemble mode, or when mechanical backaction on that mode can be neglected, the cold external microwave port gives
\begin{equation}
n_S^{(0)}
=
\frac{\gamma_{s,\mathrm{int}}}{\gamma_s}
n_{s,\mathrm{int}},
\qquad
\gamma_s=
\gamma_{s,\mathrm{ext}}+\gamma_{s,\mathrm{int}}.
\label{eq:nS_appendix}
\end{equation}
The superscript distinguishes this bare occupation from the actual superconducting population in the coupled system. When the superconducting ensemble and optical modes respond rapidly enough to be eliminated, their resonant contributions to mechanical damping are
\begin{equation}
\Gamma_{sm}=\frac{4|G_{sm}|^2}{\gamma_s},
\qquad
\Gamma_{om}(n_c)=
\frac{4|G_{om}(n_c)|^2}{\kappa},
\label{eq:reservoir_induced_rates}
\end{equation}
where $G_{om}(n_c)=g_{0,om}\sqrt{n_c}$ and $\kappa=\kappa_{\mathrm{ext}}+\kappa_{\mathrm{int}}$. In this rate-equation limit,
\begin{equation}
\begin{aligned}
\Gamma_{\mathrm{tot}}^{(3)}
&=\Gamma_m+\Gamma_{sm}+\Gamma_{om},\quad
n_m^{(3)}
&\simeq
\frac{\Gamma_{sm}n_S^{(0)}+Q_m}
{\Gamma_{\mathrm{tot}}^{(3)}}.
\end{aligned}
\label{eq:nm_loaded_appendix}
\end{equation}
The optical reservoir contributes damping but no thermal injection under the assumed vacuum, red-sideband conditions~\cite{Zhao2025QuantumEnabledNanomechanics}. Equation~\eqref{eq:nm_loaded_appendix} requires the additional reservoir-elimination approximation, not merely weak excitation of the ensemble. If collective coupling produces appreciable hybridization, the occupations are instead obtained from the full linear Langevin equations or their steady-state covariance matrix. Weak excitation and strong coherent coupling can therefore coexist without invalidating the full linear response.

\subsection{Input-referred added noise}
At a fixed pump occupation, the normally ordered correlations of the stationary input reservoirs are
\begin{equation}
\left\langle
\hat u_\ell^\dagger[\omega]\hat u_r[\omega'] \right\rangle = 2\pi n_\ell\delta_{\ell r}\delta(\omega-\omega'),
\end{equation}
where $\hat u_\ell^\dagger[\omega]\equiv (\hat u_\ell[\omega])^\dagger$. Writing $\Xi_{a\leftarrow\ell}(\omega)$ for the scattering amplitude from reservoir $\ell$ to the external optical output, the conversion efficiency and output noise spectrum are
\begin{align}\label{eq:noise}
\eta(\omega)
&= \left|\Xi_{a\leftarrow s_{\mathrm{ext}}}(\omega)\right|^2,\quad \mathcal N_{\mathrm{out}}(\omega) = \sum_{\ell\ne s_{\mathrm{ext}}} \left|\Xi_{a\leftarrow\ell}(\omega)\right|^2 n_\ell .
\end{align}
The input-referred added noise is $n_{\mathrm{add}}(\omega) = \frac{\mathcal N_{\mathrm{out}}(\omega)}{\eta(\omega)}$. Vacuum inputs contribute nothing to the normally ordered noise, and no half-quantum is included~\cite{Zeuthen2020FiguresMerit}. For the three-mode serial coupling topology, division of the mechanical-noise and signal transfer coefficients gives
\begin{equation}
\begin{aligned}
n_{\mathrm{add}}^{(3)}(\omega,n_c)
={}&
\frac{\gamma_{s,\mathrm{int}}}
{\gamma_{s,\mathrm{ext}}}
n_{s,\mathrm{int}}+
\frac{|\chi_s^{-1}(\omega)|^2}
{\gamma_{s,\mathrm{ext}}|G_{sm}|^2}
Q_m(n_c),
\end{aligned}
\label{eq:nadd_full_frequency}
\end{equation}
where $\chi_s^{-1}(\omega)= \gamma_s/2-i(\omega-\delta_s)$ and $\delta_s$ is the ensemble-mode detuning in the chosen rotating frame. This expression follows from the full linear response and does not require the superconducting and optical modes to be eliminated. At the superconducting ensemble resonance, $\omega=\delta_s$, it reduces to
\begin{align}
n_{\mathrm{add}}^{(3)}
&=
\frac{1-\eta_s}{\eta_s}n_{s,\mathrm{int}}
+\frac{Q_m}{\eta_s\Gamma_{sm}}=
\frac{n_S^{(0)}}{\eta_s}
+\frac{\bar n_m}{\eta_s C_{sm}},
\label{eq:nadd_occupancy_form}
\end{align}
with $\eta_s=\frac{\gamma_{s,\mathrm{ext}}}{\gamma_s}$ and $C_{sm}(n_c)= \frac{4|G_{sm}|^2}{\gamma_s\Gamma_m(n_c)}$. Outside the rate-equation limit, $\Gamma_{sm}$ in Eq.\ \eqref{eq:reservoir_induced_rates} is shorthand for $4|G_{sm}|^2/\gamma_s$, as opposed to an independently established physical linewidth. The pump dependence of the mechanical noise enters through $Q_m(n_c)$, while the accompanying change in $\Gamma_m(n_c)$ also affects conversion efficiency. For Eq.\ \eqref{eq:hot_bath_power_laws}, the hot-bath injection scales as $\Gamma_h n_h\propto n_c^{0.91}$. Consequently, increasing optical power raises efficiency at the cost of added noise. If the conversion maximum shifts away from the superconducting ensemble resonance, Eq.\ \eqref{eq:nadd_full_frequency} must be evaluated at the actual operating frequency since a carrier noise value cannot be paired with an off-carrier peak efficiency.

For the four-mode architecture, Eq.\ \eqref{eq:noise} remains valid with the full four-mode scattering matrix and the auxiliary noise channels included. In a relaxation-only, unsaturated linear description, the auxiliary susceptibility is $\chi_q(\omega)$ of Appendix~\ref{app:efficiency_bandwidth} with $\Gamma_{2,\mathrm{aux}}\rightarrow\gamma_{1,\mathrm{aux}}/2$. Eliminating the auxiliary amplitude introduces the mechanical self-energy and associated effective damping
\begin{align}
\Sigma_{\mathrm{aux}}(\omega)
=g_{\mathrm{aux}}^2\chi_q(\omega),\quad \Gamma_{\mathrm{aux}}(\omega)
=2\operatorname{Re}\Sigma_{\mathrm{aux}}(\omega)
=g_{\mathrm{aux}}^2\gamma_{1,\mathrm{aux}}
|\chi_q(\omega)|^2.
\end{align}
For an auxiliary reservoir of effective occupation
$n_{\mathrm{aux}}$, the added noise becomes
\begin{equation}
\begin{aligned}
n_{\mathrm{add}}^{(4)}(\omega)
={}&n_{\mathrm{add}}^{(3)}(\omega)+
\frac{|\chi_s^{-1}(\omega)|^2}
{\gamma_{s,\mathrm{ext}}|G_{sm}|^2}
\Gamma_{\mathrm{aux}}(\omega)n_{\mathrm{aux}},
\end{aligned}
\label{eq:nadd_auxiliary}
\end{equation}
under these linear reservoir assumptions. If the auxiliary susceptibility is slowly varying across the conversion band and all three coupled subsystems can be treated as reservoirs for the mechanics, the corresponding occupation estimate is
\begin{equation}
n_m^{(4)}
\simeq
\frac{
\Gamma_{sm}n_S^{(0)}
+Q_m+\Gamma_{\mathrm{aux}}n_{\mathrm{aux}}
}{
\Gamma_m+\Gamma_{sm}+\Gamma_{om}+\Gamma_{\mathrm{aux}}
},
\label{eq:nm4_rate}
\end{equation}
with the effective rates evaluated at the operating resonance. A cold auxiliary reservoir contributes no additional thermal photons, but its self-energy can still modify conversion efficiency and spectral structure. Compensating the frequency shift at one operating frequency does not remove the frequency dependence of $\Sigma_{\mathrm{aux}}(\omega)$ and therefore does not generally make the three- and four-mode spectra identical.

\section{Residual and collected-mode retrieval}
\label{app:robustness_retrieval}
This appendix specifies the calculations supporting Fig.~\ref{fig:robustness}. The selective-loading calculation retains a prescribed mechanical anharmonicity, whereas the optical-output calculations retain the bare auxiliary and its local dissipation. The qubits are represented in the full individual-qubit Hilbert space.

\subsection{Uniform and nonuniform pair interactions}
\label{app:zz_model}

Residual conditional frequency shifts are modeled by
\begin{equation}
 \frac{\hat H_{ZZ}}{\hbar} =\sum_{i<j}\zeta_{ij}\hat n_i\hat n_j,
 \qquad \hat n_i=\hat\sigma_i^+\hat\sigma_i^-.
\end{equation}
For uniform coupling, $\zeta_{ij}=\bar\zeta$,
\begin{equation}
 \frac{\hat H_{ZZ}}{\hbar} =\frac{\bar\zeta}{2}\hat N_q(\hat N_q-1),
 \qquad \hat N_q=\sum_i\hat n_i.
\end{equation}
The shift of the $k\rightarrow k-1$ transition is $\bar\zeta(k-1)$. The retuned calculation therefore adds $\hat H_{\mathrm{comp}}/\hbar=-\bar\zeta(k-1)\hat N_q$, without changing the pulse envelope or duration. After this compensation, the next unwanted loading step is detuned by $\alpha_m+\bar\zeta$. For example, in the lossless $N=4$, $k=2$ sector, the basis $\{|D_2^{4},0_m\rangle, |D_1^{4},1_m\rangle,|D_0^{4},2_m\rangle\}$ gives, up to a common energy,
\begin{equation}
 \frac{\hat H_{\mathrm{load}}(t)}{\hbar} =\begin{pmatrix}
 0&\sqrt{6}\,g_{\mathrm{sm}}(t)&0\\
 \sqrt{6}\,g_{\mathrm{sm}}(t)&0&\sqrt{8}\,g_{\mathrm{sm}}(t)\\
 0&\sqrt{8}\,g_{\mathrm{sm}}(t)&\alpha_m+\bar\zeta
 \end{pmatrix}.
\end{equation}

\begin{figure}[h]
    \centering
    \includegraphics[width=\linewidth]{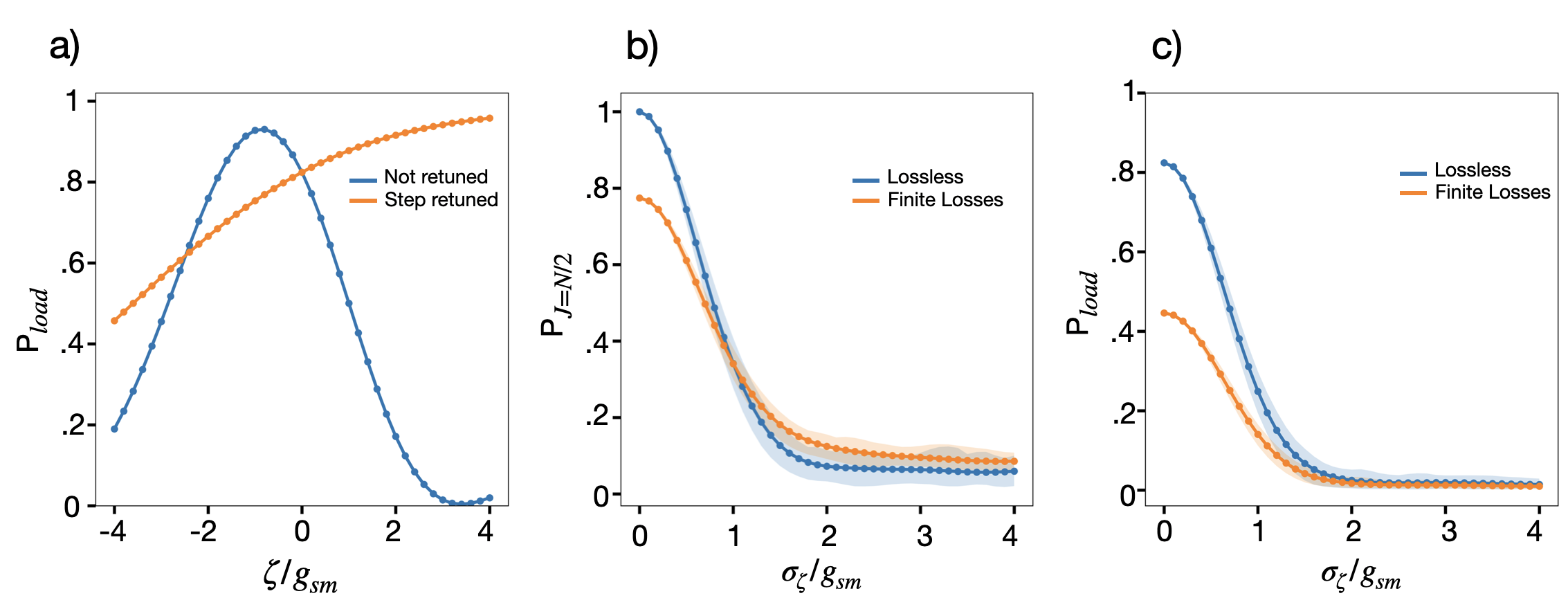}
    \caption{Sensitivity to residual pair interactions with $g_{\mathrm{sm}}/(2\pi)=10~\mathrm{kHz}$, $T_1=100~\mu\mathrm{s}$, and $T_\phi=200~\mu\mathrm{s}$. \textbf{(a)} Selected mechanical-loading probability for $N=6$, $k=3$ versus the uniform pair interaction $\bar\zeta/g_{\mathrm{sm}}$ with and without selected step retuning. \textbf{(b)} Population $P_{\mathrm{sym}}=P_{J=N/2}$ remaining in the fully symmetric qubit subspace at the same time, summed over all Dicke excitation numbers and all states of the remaining subsystems. The symmetry measure therefore includes states in which the intended single-phonon transfer has not occurred. \textbf{(c)} Probability $P_{\mathrm{load}}$ of preparing the selected loading target $\lvert D_{k-1}^{N},1_m,0_o,g_{\mathrm{aux}}\rangle$ at the end of the loading pulse. Curves show lossless evolution and evolution with local relaxation, pure dephasing, and the cold mechanical bath. Pair patterns have zero sample mean and are held fixed during each pulse with shading denoting the 10th--90th realization percentiles}
    \label{fig:ZZ_loading}
\end{figure}

For nonuniform interactions, each realization is generated from independent standard-normal draws $x_{ij}$ for the $M=N(N-1)/2$ unordered pairs. We construct
\begin{align}
 \bar x&=\frac{1}{M}\sum_{i<j}x_{ij},\quad
 s_x^2=\frac{1}{M}\sum_{i<j}(x_{ij}-\bar x)^2,\\
 \zeta_{ij}&=\bar\zeta+ \sigma_\zeta\frac{x_{ij}-\bar x}{s_x},
 \qquad \zeta_{ji}=\zeta_{ij},\quad \zeta_{ii}=0.
\end{align}
Thus each realization has exactly the specified pair mean and RMS spread. The patterns are static during a pulse and are reused across disorder strengths and loss settings. The nonuniform sweep uses $\bar\zeta=0$. We report the mean final selected-state probability and the 10th--90th realization percentiles, rather than confidence intervals of the mean. Figure~\ref{fig:ZZ_loading} uses $N=6$ and $25$ realizations per disorder strength, and Fig.~\ref{fig:robustness} uses 24. For both, $\sigma_\zeta/g_{\mathrm{sm}}$ is sampled from 0 to 4 in steps of 0.1. The supporting symmetry diagnostic is
\begin{equation}
 P_{\mathrm{sym}}= \operatorname{Tr}\!\left[
 \left(\sum_{q=0}^N|D_q^{N}\rangle\langle D_q^{N}|
 \otimes I_{\mathrm{other}}\right)\rho\right].
\end{equation}
It includes population in the ensemble ground state and is not a success probability. Relaxation can populate symmetric lower Dicke states even when the intended joint target is not reached.

Figure~\ref{fig:ZZ_loading} shows how pair interactions affect selective loading near half excitation. These calculations use $g_{\mathrm{sm}}/(2\pi)=10~\mathrm{kHz}$, $T_1=100~\mu\mathrm{s}$, and $T_\phi=200~\mu\mathrm{s}$, chosen to examine the balance between transfer speed, decoherence, and finite anharmonicity. For the loading calculation, we retain the prescribed mechanical anharmonicity $\alpha_m/(2\pi)=100~\mathrm{kHz}$ and the resonant $\sin^2$ pulse used in Fig.~\ref{fig:auxiliary_loss}. For the $N=4$, $k=2$ ensemble of Fig.~\ref{fig:auxiliary_loss}, the larger coupling shortens the loading duration to $20.4~\mu\mathrm{s}$ and gives $P_{\mathrm{load}}=P(D_1^{4},1_m)\simeq0.56$ with local relaxation, pure dephasing, and mechanical thermalization active. Doubling the coupling further shortens the pulse but lowers the selected-state probability at the same anharmonicity. Thus, reducing decoherence exposure by accelerating the exchange is beneficial only while the pulse remains sufficiently selective. This parameter point illustrates the tradeoff and is not a device prediction. For uniform pair coupling, $\zeta_{ij}=\bar\zeta$ [Fig.~\ref{fig:ZZ_loading}(a)], the interaction preserves the symmetric Dicke manifold and shifts the selected $k\rightarrow k-1$ transition by $\bar\zeta(k-1)$. Compensating this common shift restores the first transition to resonance, but the next unwanted loading step is then detuned by $\alpha_m+\bar\zeta$, so retuning helps for $\bar\zeta>0$ and reduces selectivity for $\bar\zeta<0$. Uniform interactions therefore modify both the resonance condition and the separation of neighboring transitions, rather than acting as an irreversible decay channel. Pair-to-pair variation instead produces unequal phase evolution among the components of a Dicke state and cannot be removed by a single common frequency correction. Figure~\ref{fig:ZZ_loading}(c) isolates this nonuniform contribution using static, zero-mean pair-interaction patterns with RMS strength $\sigma_\zeta$. The pulse is unchanged throughout the sweep, and absolute loading probabilities are compared with and without the local and mechanical losses. Increasing the interaction spread from zero to $\sigma_\zeta=g_{\mathrm{sm}}$ reduces the mean loading probability from approximately $0.82$ to $0.25$ in the lossless calculation and from approximately $0.45$ to $0.14$ with dissipation. The shaded intervals give the 10th--90th percentiles across static realizations. The reduction already present without losses distinguishes interaction-induced distortion of collective loading from the finite probability lost to environmental decoherence.

\subsection{Pump-dependent relaxation and optical extraction}
\label{app:finite_gate_retrieval}

In the pump-occupation sweep of Figs.~\ref{fig:robustness}(e) and \ref{fig:robustness}(f), the local ensemble collapse operators are
\begin{equation}
 \hat L_{\downarrow,i} =\sqrt{T_1^{-1}+\Gamma_\downarrow}\;\hat\sigma_i^-,
 \qquad
 \hat L_{\phi,i}=\sqrt{\frac{T_\phi^{-1}+\Gamma_\phi}{2}}\;\hat\sigma_i^z,
 \qquad
 \hat L_{\uparrow,i}=\sqrt{\Gamma_\uparrow}\;\hat\sigma_i^+,
 \label{eq:appendix_pump_relaxation}
\end{equation}
where $T_1=100~\mu\mathrm{s}$ and $T_\phi=200~\mu\mathrm{s}$ are the intrinsic lifetimes and $(\Gamma_\downarrow,\Gamma_\phi,\Gamma_\uparrow)=(0.005,\,0.005,\,0.0005)(n_c/100)~\mu\mathrm{s}^{-1}$ are the added pump-induced rates. Each curve in Figs.~\ref{fig:robustness}(e) and \ref{fig:robustness}(f) switches on one added rate, and the combined curve includes all three. Independent local channels are retained for every qubit. Pump-induced auxiliary decoherence is neglected. The auxiliary's dark relaxation and dephasing and the cold mechanical reservoir remain active. The mechanical mode is initially in vacuum, although its bath has occupation $\bar n_f=0.1$. The optically induced mechanical hot bath is omitted in this sweep. The ensemble--mechanical exchange, bare auxiliary interaction, and constant optical pump remain active throughout the collection gate. Within the resonant red-sideband rotating-wave model, elimination of the optical cavity gives
\begin{equation}
 \Gamma_{\mathrm{om}}=\frac{4|G_{\mathrm{om}}|^2}{\kappa},
 \qquad
 \hat L_{\mathrm{out}}=
 \sqrt{\frac{\kappa_{\mathrm{ext}}}{\kappa}
 \Gamma_{\mathrm{om}}}\;\hat b,
 \qquad
 \hat L_{\mathrm{int}}=
 \sqrt{\frac{\kappa_{\mathrm{int}}}{\kappa}
 \Gamma_{\mathrm{om}}}\;\hat b.
\end{equation}
This reduction assumes that the optical response is effectively flat over the relevant dressed mechanical transitions and control bandwidths. 
These channels replace the explicit optical cavity and its interaction, rather than being added to it. A fixed optical phase in $\hat L_{\mathrm{out}}$ is absorbed into the receiver convention. For $j=0$ and $j=3$, the initial state is $|D_j^{6},0_m,g_{\mathrm{aux}}\rangle$, and
\begin{equation}
 C_j=\int_0^{T_{\mathrm{gate}}}
 \operatorname{Tr}\!\left[
 \hat L_{\mathrm{out}}^\dagger\hat L_{\mathrm{out}}
 \rho_j(t)\right]dt,
 \qquad T_{\mathrm{gate}}=150~\mu\mathrm{s}.
\end{equation}
Both preparations use the same controls and baths. The efficiency and added noise are then given by Eq.\ \eqref{Eq32}. These counts include all useful output modes within the gate and are not occupations of a single selected temporal mode. Even with pump-induced excitation and absorption heating switched off, the cold mechanical bath gives a nonzero $C_0$.

\section{Simulation parameters and numerical settings} \label{app:parameters}

Table~\ref{tab:model_parameters} summarizes the model parameters, parameter sweeps, and numerical settings. Experimental reference values are distinguished from assumed design parameters and prescribed benchmark values. Coherent coefficients are stored in $\mathrm{rad}/\mu\mathrm{s}$ and Lindblad decay constants in $\mu\mathrm{s}^{-1}$. 

\begin{table*}[h]
\centering
\caption{Nominal model parameters and specified sweep ranges for the collective-exchange, linear-conversion, and nonlinear-transfer calculations. Superscript letters identify literature sources and assumed values in the notes below.}
\label{tab:model_parameters} 

\small
\setlength{\tabcolsep}{7pt}
\renewcommand{\arraystretch}{1.12}

\begin{tabular}{p{0.42\textwidth} p{0.42\textwidth}}
\toprule
\textbf{Definition} & \textbf{Value} \\
\midrule

\multicolumn{2}{@{}l}{
    \textbf{Couplings and optical pump}
}\\
\addlinespace[2pt]

Single-qubit coupling, $g_{\mathrm{sm}}/(2\pi)$
& $3.81~\mathrm{kHz}$ \\

Single-photon optomechanical coupling,
$g_{0,\mathrm{om}}/(2\pi)$
& $343~\mathrm{kHz}$\textsuperscript{(a)} \\

Intracavity pump occupation, $n_c$
& $100$ (nominal) \\

Pump-enhanced coupling, $G_{\mathrm{om}}/(2\pi)$
& $3.43~\mathrm{MHz}$ at $n_c=100$ \\

\addlinespace[4pt]
\multicolumn{2}{@{}l}{
    \textbf{Damping and coherence}
}\\
\addlinespace[2pt]

Optical linewidths,
$(\kappa_{\mathrm{ext}},\kappa_{\mathrm{int}})/(2\pi)$
& $(1.350,\,0.404)~\mathrm{GHz}$\textsuperscript{(a)} \\

Total optical linewidth, $\kappa/(2\pi)$
& $1.754~\mathrm{GHz}$ \\

Optical extraction fraction,
$\eta_o=\kappa_{\mathrm{ext}}/\kappa$
& $0.7697$ \\

Cold mechanical damping, $\Gamma_f/(2\pi)$
& $0.892~\mathrm{kHz}$\textsuperscript{(a)} \\

Cold mechanical bath occupation, $n_f$
& $0.1$ \\

Ensemble relaxation and pure-dephasing times,
$(T_1,T_\phi)$
& $(20,\,100)~\mu\mathrm{s}$ \\

Auxiliary relaxation and pure-dephasing times,
$(T_{1,a},T_{\phi,a})$
& $(20,\,100)~\mu\mathrm{s}$ \\

Coherence times, $T_2,T_{2,a}$
& $28.57~\mu\mathrm{s}$ \\

\addlinespace[4pt]
\midrule
\multicolumn{2}{@{}l}{
    \textbf{Explicit auxiliary model}
}\\
\addlinespace[2pt]

Auxiliary--mechanical coupling, $g_{\mathrm{aux}}/(2\pi)$
& $1~\mathrm{MHz}$\textsuperscript{(c)} \\
Auxiliary--mechanical detuning, $\Delta_{\mathrm{aux}}/(2\pi)$
& $1.9565~\mathrm{MHz}$\textsuperscript{(c)}\\
First-transition pull, $\chi_{\mathrm{aux}}/(2\pi)=(\omega_m-\omega_{0,1})/(2\pi)$
& $420.7~\mathrm{kHz}$ \\
\addlinespace[4pt]
\multicolumn{2}{@{}l}{
    \textbf {Prescribed-Kerr benchmark}
}\\
\addlinespace[2pt]

Kerr coefficient, $K_m/(2\pi)$
& $-50~\mathrm{kHz}$ \\

Mechanical anharmonicity,
$\alpha_m/(2\pi)=-2K_m/(2\pi)$
& $100~\mathrm{kHz}$ \\

\addlinespace[4pt]
\midrule
\multicolumn{2}{@{}l}{
    \textbf{Protocols, sweeps, and numerical settings}
}\\
\addlinespace[2pt]

Loading-time comparison
& $N=2$--$64$, $k=1$ and $k=N/2$ \\

Microscopic couplings in the loading-time comparison,
$g_{\mathrm{sm}}/(2\pi)$
& $3.81,\,100,\,1000~\mathrm{kHz}$ \\

Linear-conversion ensemble sizes
& $N=4,\,24,\,64$ (efficiency sweeps and spectra) \\

Nonlinear-transfer initial state
& $N=4$, $k=2$, mechanics and optics in vacuum \\

Mechanical-damping sweep
& $\Gamma_m=10^{-5}$--$1~\mu\mathrm{s}^{-1}$\\

Bath-occupation sweep
& $\bar n_{\mathrm{th}}=0,0.1,1$ \\
Separate qubit-coherence sweeps
& $T_1,T_\phi=10$--$10^5~\mu\mathrm{s}$ \\

Static frequency-disorder width
& $\sigma_\omega/g_{\mathrm{sm}}=0$--$4$ \\

\addlinespace[4pt]
\multicolumn{2}{@{}l}{
    \textbf{Reference optical-absorption heating model}
}\\
\addlinespace[2pt]

Hot-bath damping, $\Gamma_h(n_c)/(2\pi)$
& $(1.1~\mathrm{kHz})\,n_c^{0.61}$\textsuperscript{(b)} \\

Hot-bath occupation, $n_h(n_c)$
& $1.1\,n_c^{0.30}$\textsuperscript{(b)} \\

\bottomrule
\addlinespace[4pt]
\multicolumn{2}{@{}l}{
    \textbf{Residual interactions and optical retrieval}
}\\

Uniform pair-interaction sweep
& $\bar\zeta/g_{\mathrm{sm}}=-4$--$4$\\
Pair-disorder sweep
& $\bar\zeta=0$,
$\sigma_\zeta/g_{\mathrm{sm}}=0$--$4$\\

Optical collection interval
& $T_{\mathrm{gate}}=150~\mu\mathrm{s}$\\

Retrieval and pair-interaction calculations (Figs.~\ref{fig:robustness} and \ref{fig:ZZ_loading})
& $N=6$, $k=3$; $g_{\mathrm{sm}}/(2\pi)=10~\mathrm{kHz}$; $(T_1,T_\phi)=(100,\,200)~\mu\mathrm{s}$\\

Pump-occupation sweep
& $n_c=1$--$500$\\

Added pump-induced rates, $(\Gamma_\downarrow,\Gamma_\phi,\Gamma_\uparrow)$
& $(0.005,\,0.005,\,0.0005)(n_c/100)~\mu\mathrm{s}^{-1}$\\
\bottomrule
\end{tabular}

\vspace{4pt}
\begin{minipage}{\textwidth}
\footnotesize
\raggedright

\textsuperscript{(a)}\,H. Zhao \textit{et al.},
\textit{Nature Nanotechnology} \textbf{20}, 602--608 (2025).

\smallskip
\textsuperscript{(b)}\,H. Ren \textit{et al.},
\textit{Nature Communications} \textbf{11}, 3373 (2020).

\smallskip
\textsuperscript{(c)}\,Assumed comparison parameters giving approximately $100~\mathrm{kHz}$ anharmonicity from the exact two-level Jaynes--Cummings spectrum. The first-transition pull $\chi_{\mathrm{aux}}$ is calculated from the same spectrum. Used in the retrieval calculations of Fig.~\ref{fig:robustness}, not used in the prescribed-Kerr simulations.

\end{minipage}
\end{table*}

\newpage

\bibliography{references}
\end{document}